%% file: main_revised_arXiv.tex
\documentclass[aps,10pt,reprint,groupedaddress,superscriptaddress, prx]{revtex4-2}
\usepackage{amssymb}
\usepackage{graphicx}
\usepackage{booktabs}
\usepackage{amsmath}
\usepackage{multirow}
\usepackage{blkarray}
\usepackage{nicematrix}
\usepackage{tikz}
\usepackage{mathrsfs}
\usepackage[bookmarks=true,colorlinks,citecolor=blue,urlcolor=blue]{hyperref}
\usepackage{braket}
\usepackage{babel}
\usepackage{blindtext}
\usepackage{soul}
\usepackage[compat=1.1.0]{tikz-feynman}
\usepackage[normalem]{ulem}
\usepackage{physics}

\usepackage{mathtools}

\definecolor{mypurp}{rgb}{0.35, 0, 0.7}

\usepackage{amsthm}
\usepackage{txfonts}

\theoremstyle{definition}

\newcommand{\tTC}{\text{TC}}
\newcommand{\tI}{\text{Ising}}
\newcommand{\tTSB}{\text{TSB}}

\newcolumntype{C}[1]{>{\centering\arraybackslash}p{#1}}

\begin{document}

\def\papertitle{Tensor-network study of the roughening transition in a (2
+ 1)D $\mathbb{Z}_2$ lattice gauge theory with matter}

\newcommand{\TUM}{\affiliation{Technical University of Munich, TUM School of Natural Sciences, Physics Department, 85748 Garching, Germany}}
\newcommand{\MCQST}{\affiliation{Munich Center for Quantum Science and Technology (MCQST), Schellingstr. 4, 80799 M{\"u}nchen, Germany}}
\author{Wen-Tao Xu} \TUM \MCQST
\author{Michael Knap} \TUM \MCQST
\author{Frank Pollmann} \TUM \MCQST

\title{\papertitle}
\begin{abstract}
Within the confined phase of (2+1)D lattice gauge theories a roughening transition arises between a weakly confined regime with floppy string excitations and a strongly confined regime with stiff string excitations. In this work, we use an infinite Density Matrix Renormalization Group (iDMRG) algorithm to quantitatively characterize the properties of confined strings.
To this end, we stabilize the state with a string excitation by 't~Hooft loop operators. While for zero gauge-matter coupling we can use bare 't~Hooft loop operators to do so, for finite gauge-matter coupling we have to transform them to  emergent ones, which we achieve with an adiabatic protocol. By analyzing the scaling of both a novel order parameter and the entanglement entropy, our approach allows us to accurately determine the roughening transition, even at finite gauge-matter coupling.
\end{abstract}

\maketitle

\begin{figure}[tbp]
\centering
\includegraphics[width=0.99\linewidth]{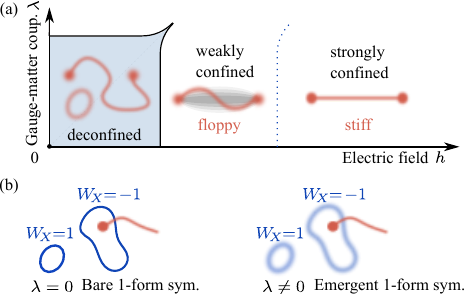}
\caption{\textbf{Confinement of electric charges in a gauge theory.} (a) Schematic phase diagram of the $\mathbb{Z}_2$ lattice gauge theory with Ising matter~\cite{Fradkin-Shenker}. When the electric field is weak, the electric charges (red dots) are deconfined and the electric string (red lines) between them moves freely without energy cost. With the electric field increasing, there is a 
phase transition to the confined phase where the electric charges are bound. Within the confined phase a roughening transition arises (dotted line), which we show to bend to the right as the gauge-matter coupling increases. At intermediate electric field, the width of the electric flux tube diverges logarithmically with the distance between two electric charges and the confined electric string is floppy. For strong electric field the width of the electric flux tube is constant and the confined electric string is stiff.  (b) When the gauge-matter coupling $\lambda$ is zero, the 't Hooft loop 
$W_X$, measuring the electric charges, is a bare 1-form symmetry and commutes with the Hamiltonian. For finite gauge-matter coupling $\lambda$ the 't Hooft loop $W_X$ becomes an emergent 1-form symmetry.}
\label{Fig:intro}
\end{figure}

\textbf{Introduction.} Confinement is a fundamental phenomenon in physics~\cite{Wilson_confinemnt_1974} in which particles are bound together and form new composite particles.
In condensed matter physics, (1+1)D spin chains represent an illustrative example for confinement, where mesonic bound states are formed by associating an energy cost with domain-wall separation~\cite{McCoy1978, Rutkevich2008}, leading to exotic non-equilibrium dynamics~\cite{confinement_quench_2017, Mazza2019, Liu2019, Lerose2020, zhu_2023,Prethermal_2022}. Confinement in (1+1)D spin chains has been experimentally investigated with quantum simulators~\cite{Simon2011, Tan2021, Knolle_2021, de2024, Mildenberger2025}. The situation is fundamentally distinct in higher dimensions, as the string that confines the charges also has transverse degrees of freedom  (Fig.~\ref{Fig:intro}a).
Interestingly, depending on the strength of the confining potential, the width of the flux tube connecting the two charges is qualitatively distinct~\cite{LUSCHER_1981_expansion, LUSCHER_1981_SSB_aspect,HASENFRATZ_1981_SOS}, which has recently been demonstrated in the string dynamics for a (2+1)D lattice gauge theory on a quantum processor~\cite{visualizing_2024}. When considering eigenstates with string excitations, the transverse fluctuations of the string are suppressed for a strong linear confinement potential, resulting in a stiff flux tube with finite thickness. In contrast, for weak linear confining potential, the transverse fluctuations are strong, so that the string is floppy and the width of the flux tube diverges logarithmically with the distance between charges~\cite{LUSCHER_1981_expansion,numerical_width_logrithmic_1996}. A roughening transition separates the two cases in the limit of infinitely long strings, which has been predicted for various lattice gauge theories~\cite{MUNSTER_1981_expansion,DROUFFE_1981,DROUFFE_1981_2}.

For a pure $\mathbb{Z}_2$ lattice gauge theory with vanishing gauge-matter coupling, the roughening transition can be understood from a duality with the Ising model~\cite{Wegner_duality_1971}. A string excitation in gauge theory leads to a domain-wall interface in the Ising model; the roughening transition of the pure $\mathbb{Z}_2$ gauge theory corresponds to the roughening of a domain-wall interface in the ferromagnetic phase of the Ising model, which has been intensively studied by analyzing the width of the interface~\cite{Fradkin_1983,Burkner_1983_MC_Ising,Matching_Ising_sos_1997,MC_Ising_interface_2003,3d_Ising_interface_Nabum_Goto_2007,Roughen_3D_classical_Ising_2019,roughening_dynamics_2024}. However, since the
roughening transition is a Berezinskii-Kosterlitz-Thouless (BKT) transition~\cite{KT_1973,Berezinskii_1971,Berezinskii_1972,HASENFRATZ_1981_SOS}, which is notoriously hard to locate numerically, a quantitative way of determining the roughening transition point is still lacking.
To study confinement in gauge theories, one detects electric charges and the string connecting them by 't Hooft loops,
which can be interpreted as a generalized loop-like 1-form symmetry~\cite{Zohar_2004,NUSSINOV_2009,High_form_Kapistin_2015,McGreevy_2023,Bhardwaj_2023}, see Fig.~\ref{Fig:intro}b. However, for finite gauge-matter coupling, the 1-form 't Hooft loop symmetry becomes an emergent symmetry~\cite{Hastings_and_Wen_2005,High_form_wen_2019,Adam_Nahum_2021,Wen_emergent_high_form_2023, Emergent_1_form_PRD_2024,Adam_Nahum_2024,QEC_1_form_2025}, and is a dressed loop operator that is not directly accessible. For the same reason, the duality to a simple Ising model breaks down, and one needs to develop novel methods to explore the roughening transition at finite gauge-matter couplings, which is a challenging problem and has not been investigated thus far.

In this work, we develop and utilize tensor network methods to study the roughening transition at finite gauge matter coupling. We use entanglement and symmetry to quantitatively characterize the roughening transition, distinct from previous qualitative observations based on the flux tube width or the so-called ``kink mass"~\cite{Kogut_1981_PRD_1,Kogut_1981_PRD_2}. The effective model of the roughening transition is (1+1)D and describes the transverse displacement of the 1D string excitation, although the lattice gauge theory is $(2+1)$D~\cite{HASENFRATZ_1981_SOS,roughening_dynamics_2024}. Because of this dimensional reduction, we consider quasi-1D lattice gauge theories on cylinder geometries and find the infinite Density Matrix Renormalization Group (iDMRG) algorithm~\cite{DMRG_1992,DMRG_1993,mcculloch_2008_iDMRG,SCHOLLWOCK_2011} to be well suited to study the roughening transition in (2+1)D gauge theories.
To access the roughening transition at finite gauge-matter coupling,
we develop an adiabatic protocol by utilizing a sequence of iDMRG simulations
and show that the floppy string is in fact stable against gauge-matter coupling. Our method provides a versatile computational tool for simulating and characterizing confinement, allowing us to locate the roughening transitions of $(2+1)$D quantum lattice gauge theories with both zero and finite gauge-matter couplings.

\begin{figure}[tbp]
\centering
\includegraphics[width=0.99\linewidth]{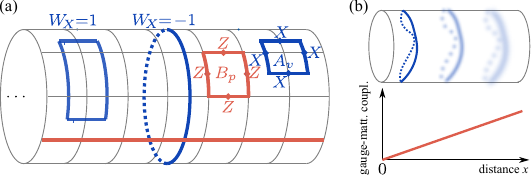}
\caption{\textbf{Simulating an electric string of the $\mathbb{Z}_2$ lattice gauge theory on a cylinder using tensor networks.} (a) Vertex operator ($A_v$), plaquette operator ($B_p$), bare 1-form 't Hooft loop operator on the dual lattice (blue loop), and electric string on the primal lattice (red line). (b) An infinite system-size DMRG (iDMRG) simulation with slowly increasing gauge-mater coupling $\lambda$ can be interpreted as a spatially inhomogeneous system in which the gauge-matter coupling slowly increases with distance $x$. This process transforms the bare 't Hooft loop operator into an emergent 't Hooft loop operator and allows us to probe the roughening transition for finite gauge matter coupling.}
\label{Fig:method}
\end{figure}

\textbf{(2+1)D lattice gauge theory.}
We consider the Fradkin-Shenker model as a $\mathbb{Z}_2$ lattice gauge theory with the $\mathbb{Z}_2$ Ising matter~\cite{Kogut_Susskind,Kogut_LGH,Fradkin-Shenker}. By choosing the unitary gauge, the Ising spin matter degrees of freedom can be discarded and the Fradkin-Shenker model is simplified to the toric code in a field~\cite{TC_multi_critical_2010,xu_2024_entanglement}:
\begin{equation}\label{eq:TC_Hamiltonian}
H=-J_E\sum_{v}A_v-J_M\sum_{p}B_p-h\sum_{e}X_e-\lambda\sum_{e} Z_e,
\end{equation}
where $A_v=\prod_{e\in v}X_e$ and $B_p=\prod_{e\in p} Z_e$ are the vertex and plaquette operators of the toric code model~\cite{kitaev_2002}, and $X_e$ and $Z_e$ are Pauli matrices on the edges $e$ of the lattice, see Fig.~\ref{Fig:method}a. Here, $J_E>0$ and $J_M>0$ control the gaps of electric and magnetic excitations, respectively, $h$ is the electric field, and $\lambda$ the gauge-matter coupling (the matter field disappeared because of the unitary gauge). A schematic phase diagram of the model is shown in Fig.~\ref{Fig:intro}a. When both $h$ and $\lambda$ are small, electric charges are deconfined. When $h$ is large and $\lambda$ is small, electric charges are confined.

For vanishing gauge-matter coupling $\lambda=0$, the model has a bare 1-form 't Hooft loop symmetry $W_X=\prod_{e\in \hat{C}}X_e$, where $\hat{C}$ is a loop along the dual lattice, i.e., $[H,W_X]=0$, see Fig.~\ref{Fig:method}a. The 1-form 't Hooft loop symmetry measures the $\mathbb{Z}_2$ electric charges surrounded by the loop $\hat{C}$. For finite gauge-matter coupling  $\lambda\neq0$, the bare 1-form 't Hooft loop symmetry is no longer a symmetry of $H$ and the electric charges are measured by an emergent 1-form 't Hooft loop symmetry~\cite{Hastings_and_Wen_2005} challenging their numerical detection, see Fig.~\ref{Fig:intro}b.

\textbf{Characterizing the roughening transition for vanishing gauge matter coupling.} We will first illustrate our numerical approach and introduce the observables for detecting the roughening transition for the pure lattice gauge theory ($\lambda =0$).
To obtain the eigenstate of $H$ in Eq.~\eqref{eq:TC_Hamiltonian} with an electric flux string, we calculate the ground state of the Hamiltonian: $\tilde{H}=H+J_W\sum_{x}W_X^{[y]}(x)$ using iDMRG~\cite{DMRG_1992,DMRG_1993,mcculloch_2008_iDMRG,SCHOLLWOCK_2011}, where $W_X^{[y]}(x)$ is a non-contractible 't Hooft loop operator wrapping in the $y$ direction whose horizontal coordinate is $x$, see Fig.~\ref{Fig:method}a. Since both $W_X^{[y]}(x)$ and $A_v$ commute with $H$, the ground state of $\tilde{H}$ does not depend on $J_W$ and $J_E$ as long as they are sufficiently large.

\begin{figure}
\centering
\includegraphics[width=0.99\linewidth]{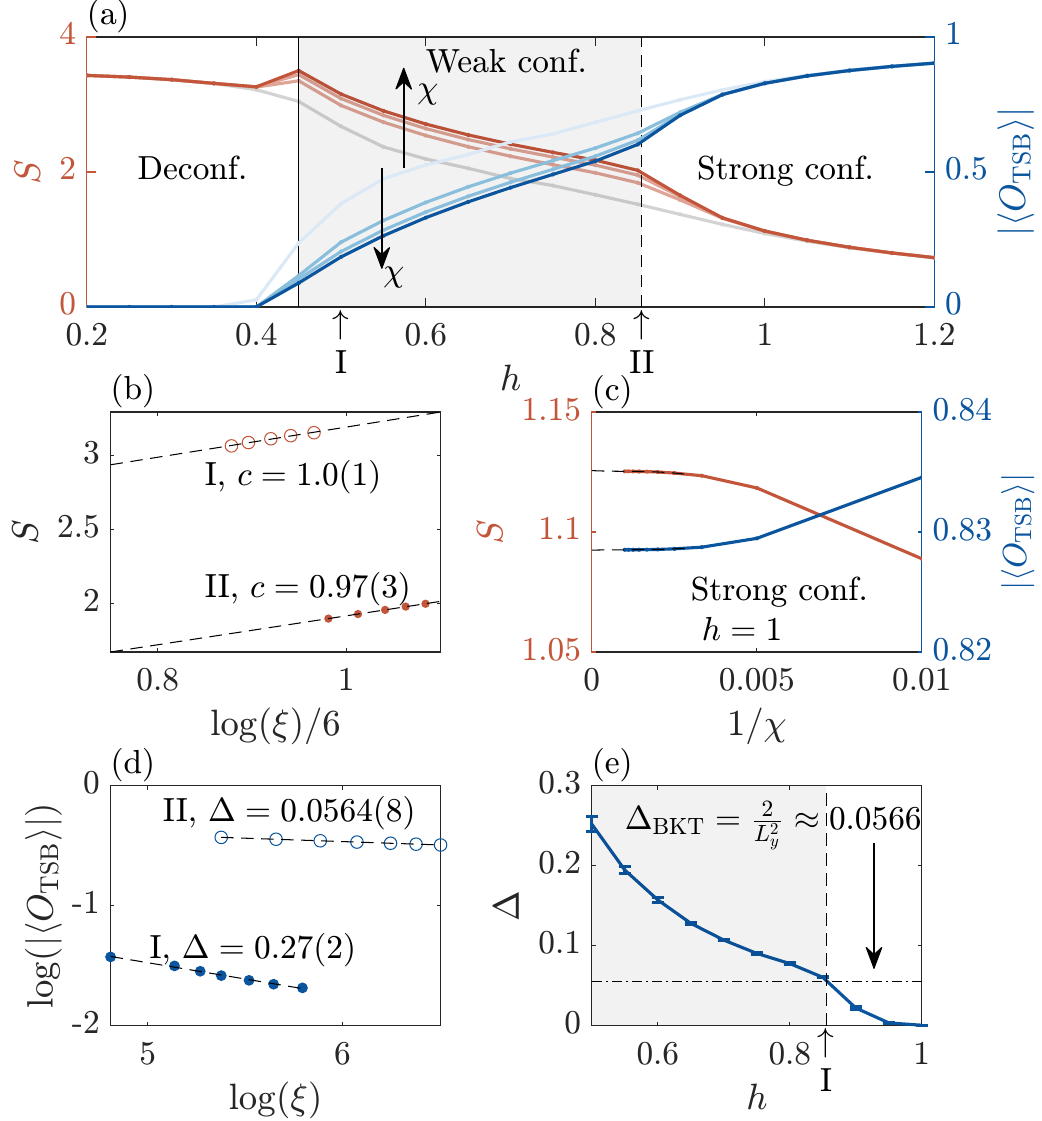}
\caption{\textbf{Roughening transition for vanishing gauge-matter coupling ($\lambda = 0$).} (a) Entanglement entropy $S$ and translational symmetry breaking order parameter $|\langle O_{\tTSB}\rangle|$ obtained from iDMRG simulations with bond dimensions $\chi=100,400,700,1000$ and $L_y=6$.
(b) The central charge $c$ extracted from the entanglement entropy as a function of the correlation length $\xi$  
at $h=0.5$ (``I'') and at the estimated BKT transition point at $h_{\text{BKT}}\approx0.855$ (``II''). (c) $S$ and $|\langle O_{\tTSB}\rangle|$ as a function of $\chi$ at $h=1$. The dashed lines represent extrapolations to $\chi\rightarrow\infty$. (d) Order parameter $|\langle O_{\tTSB}\rangle|$ as function of $\xi$. (e) The extracted scaling dimension $\Delta$ of $|\langle O_{\tTSB}\rangle|$ as a function of $h$. The dash-dot line indicates the scaling dimension $\Delta_{\text{BKT}}$ of $|\langle O_{\tTSB}\rangle|$ at the roughening (BKT) transition point. 
}
\label{Fig:1}
\end{figure}

As a first observable, we will compute the entanglement entropy $S$ of the eigenstates obtained from iDMRG to detect the different behaviors of the  strings at weak and strong confinement; see Fig.~\ref{Fig:1}a.
We consider cylinders that are infinitely long in $x$-direction and have a finite circumference $L_y=6$ in $y$-direction; results for larger systems are shown in the supplementary material~\cite{appendix}.
Analyzing the entanglement entropy $S$ of the state obtained with different bond dimensions $\chi$, we find that both for very small and very large electric field $h$, corresponding to the deconfined phase and the strongly confined regime, respectively, the entanglement $S$ converges with bond dimension $\chi$ (see Fig.~\ref{Fig:1}c for strong confinement). In contrast, for intermediate electric field strength $h$, $S$ increases with bond dimension, indicating a gapless intermediate phase. In particular, for a critical phase described by a conformal field theory,  we expect that the entanglement scales as $S=c\log(\xi)/6$, with $c$ being the central charge and $\xi$ is the correlation length extracted from the state with a bond dimension $\chi$. Performing finite entanglement scaling by extracting the entanglement entropy $S$ and the correlation correlation length $\xi$ of the states from the iDMRG simulations with various bond dimensions $\chi$~\cite{Luca_2008,Pollmann_2009}, we obtain $c\approx 1$, as shown in Fig.~\ref{Fig:1}b, which is the expected behavior for a BKT phase. The entanglement entropy thus clearly distinguishes between the strong and weak confinement. A related but qualitative observation from finite DMRG has been studied in a lattice U(1) gauge theory ~\cite{Marcello_Dalmonte_2019}.
Since at zero gauge-matter coupling, the $\mathbb{Z}_2$ gauge theory is dual to the Ising model, we demonstrate that we can also use the entanglement entropy of the Ising model to determine the weakly and strongly confined regimes~\cite{appendix}.

To quantitatively locate the position of the BKT transition and characterize the BKT phase, we construct an order parameter based on the symmetry-breaking aspect of the roughening transition~\cite{LUSCHER_1981_SSB_aspect}. On a cylinder with a circumference $L_y$, see Fig.~\ref{Fig:method}a, the Hamiltonian $H$ in Eq.~\eqref{eq:TC_Hamiltonian} has $\mathbb{Z}_{L_y}$ translational symmetry.
In the strongly confined regime, there are $L_y$ degenerate eigenstates with electric strings (the horizontal red line in Fig.~\ref{Fig:method}a) located at $y=0,1,\cdots, L_y-1$; so the translation symmetry along the $y$ direction is broken spontaneously. However, in the weakly confined regime a continuous translational symmetry emerges~\cite{LUSCHER_1981_SSB_aspect}, giving rise to a BKT phase.
\footnote{The BKT phase of the weakly confined floppy string is similar to the BKT phase of the 1D $p$-state quantum clock model with a global internal symmetry $\mathbb{Z}_p$, which has an intermediate BKT phase with emergent $U(1)$ symmetry for $p\geq 5$~\cite{TSUI_2017}, even though the Hamiltonian has a discrete symmetry.}
We can define such an order parameter $|\langle O_{\tTSB}\rangle|$ to detect the translational symmetry breaking, by requiring that $O_{\tTSB}$ satisfies $\mathcal{T}^{[y]}O_{\tTSB}\mathcal{T}^{[y]\dagger}=\exp(2i\pi /L_y)O_{\tTSB}$ with $\mathcal{T}^{[y]}$ being the translation operator along the $y$-direction. We choose
\begin{equation}
    O_{\tTSB}=\sum_{e\in \hat{C}_y}\exp[i\frac{2\pi y(e)}{L_y}]\frac{1-X_e}{2},
\end{equation}
where $y(e)=0,1,\cdots,L_y-1$ labels the $y$ coordinate of the edge $e$ and $\hat{C}_y$ is a vertical non-contractible loop on the dual lattice. The order parameter detects the spatial distribution of the electric string in the $y$ direction by considering a non-zero Fourier mode of the distribution, and captures translational symmetry breaking.
Since the roughening transition is a BKT transition accompanied by spontaneous $\mathbb{Z}_{L_y}$ symmetry breaking, the $\mathbb{Z}_{L_y}$-deformed sine-Gordon model~\cite{Wiegmann_1978,Matsuo_2006} is a well-suited effective field theory for this transition. From the $\mathbb{Z}_{L_y}$-deformed sine-Gordon model, we derive the scaling dimension of $O_{\tTSB}$, see supplemental material~\cite{appendix} and Tab.~\ref{tab}. From the scaling dimensions, we then obtain the Luttinger parameter $R$ (compactified radius of the boson field) in the BKT phase and locate the BKT transition point. For example, since the electric field drives the BKT transition, the corresponding field in the sine-Gordon model becomes marginally irrelevant with a scaling dimension $2$ at the roughening transition. Based on this, we obtain the scaling dimensions of  $O_{\tTSB}$ at the BKT transition as $2/L_y^2$.

\begin{table}
    \centering

    \begin{tabular}{C{2cm}C{2cm}C{2cm}C{2cm}}
    \toprule
          Operator &   $X_e-\langle X_e\rangle$ &  $O_{\text{TSB}}$  \\
           \midrule
        $\Delta$ &  $L_y^2/R^2$  & $1/R^2$    \\
          $\Delta_{\text{BKT}}$ &  $2$  & $2/L_y^2$     \\
          \bottomrule
    \end{tabular}
    \caption{\textbf{Scaling dimensions.} Scaling dimensions $\Delta$ in the gapless BKT phase and $\Delta_{\text{BKT}}$ at the roughening transition, where $R$ is the Luttinger parameter, or equivalently the radius of a compactified free boson conformal field theory, and $L_y$ is the cylinder circumference. The electric field $\sum_e X_e$ drives the roughening transition, $O_{\tTSB}$ is the order parameter detecting the translational symmetry breaking in the strongly confined phase.}
    \label{tab}
\end{table}

The results for $|\langle O_{\tTSB}\rangle|$ obtained from iDMRG simulations with various bond dimensions
are shown in Fig.~\ref{Fig:1}a. In the deconfined phase $|\langle O_{\tTSB}\rangle|=0$ while $|\langle O_{\tTSB}\rangle|$ converges to a non-zero value in the strongly confined regime, see inset of Fig.~\ref{Fig:1}c. For weak confinement, $|\langle O_{\tTSB}\rangle|$ slowly decreases with  bond dimension $\chi$. A finite correlation scaling of $|\langle O_{\tTSB}\rangle|$ using the correlation length $\xi$ induced by the finite bond dimensions shows that $|\langle O_{\tTSB}\rangle|$ decays algebraically to zero with $\xi$ in the weakly confining regime, see Fig.~\ref{Fig:1}d. The exponent of the algebraic decay corresponds to the scaling dimension $\Delta$ of $|\langle O_{\tTSB}\rangle|$, whose relation with the Luttinger parameter is displayed in Tab.~\ref{tab}. We plot the extracted values of $\Delta$ as a function of $h$ in Fig.~\ref{Fig:1}e~\footnote{When $h > h_{\text{BKT}}$, the correlation length $\xi(\chi \to \infty)$ is finite. However, as long as we remain in the regime where $\xi(\chi) \ll \xi(\chi \to \infty)$, we can still extract a scaling dimension from the finite-$\chi$ data.}. Since the scaling dimension of $|\langle O_{\tTSB}\rangle|$ is $2/L_y^2$ at the BKT transition, see Tab.~\ref{tab}, we locate the BKT transition at $h_{\text{BKT}}\approx 0.855$ for $L_y=6$ in Figs.~\ref{Fig:1}d and e and $h_{\text{BKT}}\approx 0.865$ for $L_y=8$~\cite{appendix}. This is consistent with the expectation that with $L_y$ increasing, $h_{\text{BKT}}$ becomes slightly larger, because the smaller cylinder circumference suppresses the transverse fluctuation of the electric string.  Note that our results do not quantitatively agree with previous estimates of $h_{\text{BKT}}\approx 0.59$ in Ref.~\cite{HASENFRATZ_1981_SOS} and $h_{\text{BKT}}\approx 0.72$ in Ref.~\cite{roughening_dynamics_2024} using the effective solid-on-solid model describing the confined string. A possible reason for the discrepancy is that the effective solid-on solid model only accounts for the transverse fluctuation of the electric string, and the longitudinal fluctuation of the electric string and the fluctuation of the electric loops are ignored, yet their effects can be significant at weak confinement. Since we directly consider the original model, the effects of these fluctuations are accounted for. For pure gauge theories, the Wilson loop operator is also commonly used to detect the transition between confined and deconfined phases~\cite{Wegner_duality_1971,Kogut_LGH}.  In the supplemental material~\cite{appendix} we show, that the decay of Wilson loop operator correlations also distinguishes strong and weak confinement.

\begin{figure}[tbp]
\centering
\includegraphics[width=0.99\linewidth]{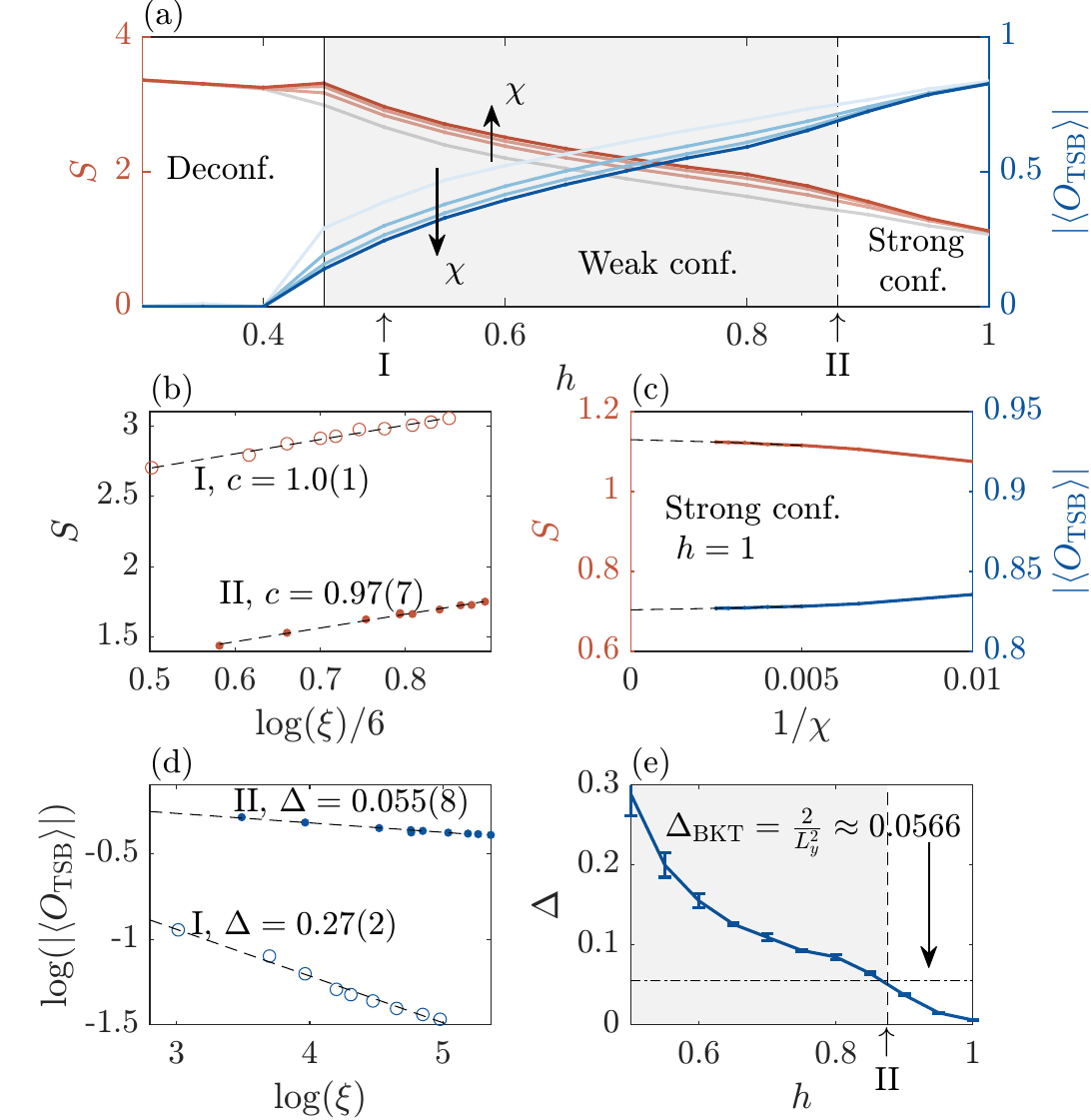}
\caption{\textbf{Roughening transition for finite gauge-matter coupling.}  (a) Entanglement entropy $S$ and translational symmetry breaking order parameter $|\langle O_{\tTSB}\rangle|$ obtained from iDMRG simulations with $\chi=100,200,\cdots,400$.
(b) The central $c$ extracted from the entanglement entropy as a function of the correlation length $\xi$
at $h=0.5$ (``I'') and at the estimated BKT transition point $h_{\text{BKT}}\approx0.875$ (``II''). (c) $S$ and $|\langle O_{\tTSB}\rangle|$ as a function of $\chi$ at $h=1$. The dashed lines represent extrapolations to $\chi\rightarrow\infty$. (d) Order parameter $|\langle O_{\tTSB}\rangle|$ as function of $\xi$.
We set $L_y=6,\lambda=0.2,J_M=1,J_E=2$. (e) The extracted scaling dimension $\Delta$ of $|\langle O_{\tTSB}\rangle|$ as a function of $h$. The dash-dot line indicates the scaling dimension $\Delta_{\text{BKT}}$ of $|\langle O_{\tTSB}\rangle|$ at the roughening (BKT) transition point. }
\label{Fig:2}
\end{figure}

\textbf{Roughening transition for finite gauge-matter coupling.} When the gauge-matter coupling $\lambda$ is non-zero, the bare 1-form 't Hooft loop operator is no longer a symmetry of the Hamiltonian $H$. Instead there exists an emergent 1-form 't Hooft loop symmetry which depends on the details of the model. It is difficult to find the explicit expression of an emergent 1-form symmetry.
We thus develop an adiabatic protocol to approach the case of the emergent 1-form 't Hooft loop operator using iDMRG. For this, we first obtain the eigenstate with the electric string at zero gauge-matter coupling and then slowly ramp up the gauge-matter coupling while targeting the eigenstate with the electric string. We mimic the adiabatic evolution by sequential iDMRG simulations~\cite{Pollmann_FCI_2015,tenpy}; a technique which has been used to simulate charge pumping through adiabatic flux insertion in topological states~\cite{Zaletel_2014}. The basic idea is that we consider a path given by a sequence of $\{\lambda_0,\cdots,\lambda_{i},\cdots\}$ with $\lambda_0=0$ and a small interval $\epsilon=\lambda_{i+1}-\lambda_{i}$. For $\lambda_0$, we calculate the eigenstate with the confined string using the Hamiltonian $\tilde{H}=H+J_W\sum_{x}W_X^{[y]}(x)$. Then, for $\lambda_i$ with $i>0$, we run  iDMRG by initializing the iDMRG environment using the converged iDMRG environment at $\lambda_{i-1}$ and the Hamiltonian $H(J_E,J_M,h,\lambda=\lambda_i)$. The technical details of the sequential iDMRG simulations are shown in the supplemental material~\cite{appendix}. Since in iDMRG the system size effectively grows with the number of sweeps, the gauge-matter coupling is slowly increasing in $x$ direction. When pulling through a 't Hooft loop operator on such a system, it changes from the bare 1-form symmetry to an emergent 1-form symmetry, see Fig.~\ref{Fig:method}b. By imposing the electric string using the bare 't Hooft loop operator  at $\lambda=0$, we obtain the eigenstate with the electric string at $\lambda\neq 0$. We can track the eigenstate of the effective iDMRG Hamiltonian with the maximum overlap of the eigenstate from the previous sweep, in analogy with the DMRG-X method~\cite{DMRG_X_2016,DMRG_X_2017}, which further improves stability of the approach.

To investigate the robustness of the floppy string at finite gauge-matter coupling, we calculate the eigenstate with an electric string for $J_E=2,J_M=1,\lambda=0.2$ and tune $h$ 
using the sequential iDMRG simulations outlined above.
The entanglement entropy $S$ and the translational symmetry breaking order parameter $|\langle O_{\tTSB}\rangle|$ confirm the existence of the floppy string for intermediate electric field $h$; see Fig.~\ref{Fig:2}a. The central charge $c$ extracted from the states of the sequential iDMRG simulations is close to $1$ in the weakly confined regime; see Fig.~\ref{Fig:2}b, confirming further that the weakly confined regime is robust against perturbations which explicitly break the exact 1-form 't Hooft loop symmetry and that it is still a critical BKT phase. In Fig.~\ref{Fig:2}c, at $h = 1$, both the entanglement entropy and the order parameter saturate to finite values as $\chi \rightarrow \infty$, indicating that the electric flux tube is strongly confined. In Figs.~\ref{Fig:2}d and e,  we extract the scaling dimension of $|\langle O_{\tTSB}\rangle |$ as a function of $h$, yielding $h_{\text{BKT}}(\lambda=0.2)\approx0.875$, which is slightly larger than $h_{\text{BKT}}(\lambda=0)\approx0.855$ at zero gauge-matter coupling. For $L_y=8$, we find $h_{\text{BKT}}(\lambda=0.2)\approx0.88$; see supplemental material~\cite{appendix}.

\textbf{Conclusion and Outlook.} We developed and utilized tensor network based algorithms to characterize the roughening transition between a weakly and a strongly confined string excitation.
We showed that sequential iDMRG simulations can be used to investigate the confinement at non-zero gauge-matter coupling, from which we find that the weakly confined regime is stable against the non-zero gauge-matter coupling. This enabled us to explore the dependence of the roughening transition point on the gauge-matter coupling. For gauge theories in continuous space it has been argued that the confined flux tube is always rough at any finite field $h$~\cite{LUSCHER_1981_SSB_aspect}. This suggests that in our lattice model the roughening transition point $h_{\text{BKT}}$ increases with the gauge-matter coupling $\lambda$, as the width of the emergent 1-form symmetry grows with $\lambda$, effectively reducing the lattice spacing. At a critical value of the gauge-matter coupling, the emergent 1-form ’t Hooft loop symmetry ceases to exist~\cite{Adam_Nahum_2021,QEC_1_form_2025}. Near but below this threshold, the width of the 1-form symmetry is expected to become large pushing the roughening transition point to very large fields. 

Our method can be applied to a large class of lattice gauge theories, as long as the relation between the emergent ’t Hooft loop operator and the exact ’t Hooft loop is known across the parameter space. For example, we investigate the $\mathbb{Z}_3$ gauge theory in the supplemental material~\cite{appendix}. In future work, it will be exciting to study the weakly confined string connecting non-Abelian anyons~\cite{Xu_2020,Xu_2021,Xu_2022}. Moreover, we emphasize that in a lattice gauge theory the roughening transition of the confined string depends on how the string is put on the lattice. For instance, when considering a zigzag string along the diagonal direction of a pure $\mathbb{Z}_2$ lattice gauge theory on a square lattice, there is no roughening string transition as Manhattan distance preserving moves do not cost energy~\cite{visualizing_2024, borla_2025}.
By performing iDMRG on a helical lattice, one can interpolate between a zigzag string and a straight string which would be interesting to explore as well~\cite{Kogut_1981_PRD_1,Kogut_1981_PRD_2}. Furthermore, it has been predicted that the width of the flux tube at the bulk topological phase transition toward deconfinement may be modified~\cite{Rough_at_QCP_2009,Rough_at_QCP_2010}. Thus, a challenging problem is to extend our approach to study the electric string in the vicinity of the bulk topological transition, for instance with projected entangled pair states~\cite{Xu_huang_2024,xu_FM_2024}.

\textbf{Acknowledgements.} We thank Marcello Dalmonte, Torsten Zache and Rui-Zhen Huang for insightful discussions. Calculations were performed using the TeNPy Library~\cite{tenpy}. We acknowledge support from the Deutsche Forschungsgemeinschaft (DFG, German Research Foundation) under Germany’s Excellence Strategy–EXC–2111–390814868, TRR 360 – 492547816 and DFG grants No. KN1254/1-2, KN1254/2-1, FOR 5522
(project-id 499180199), the European Research Council (ERC) under the European Union’s Horizon 2020 research and innovation programme (grant agreement No 851161), the European Union (grant agreement No 101169765), as well as the Munich Quantum Valley, which is supported by the Bavarian state government with funds from the Hightech Agenda Bayern Plus.

\textbf{Data availability.}
Data and codes are available upon reasonable request on Zenodo~\cite{zenodo}.

\clearpage

\input{Appendix}

\bibliography{ref_revised_v3.bib}
\end{document}

%% file: Appendix.tex
\onecolumngrid
\setcounter{equation}{0}
\setcounter{figure}{0}
\setcounter{table}{0}
\setcounter{page}{1}

\renewcommand{\theequation}{S\arabic{equation}}
\renewcommand{\thefigure}{S\arabic{figure}}
\renewcommand{\thetable}{S\arabic{table}}

\begin{center}
    \textbf{\large Supplemental Material: \\ \papertitle}
\end{center}

\twocolumngrid

\begin{figure*}
\centering
\includegraphics[scale=1]{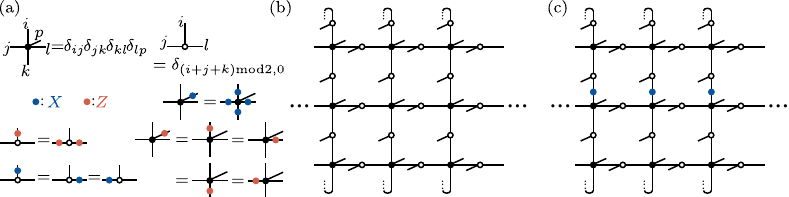}
\caption{\textbf{The tensor network operators of the duality transformations between systems on an infinite long cylinder.} (a) The tensors generating the tensor network operators and the symmetries of these tensors. The 5-leg tensor is a $\delta$-tensor: its entries are 1 when all indices are equal and 0 otherwise. Applying an $X$ operator to the physical leg of the 5-leg tensor is equivalent to applying $X$ operators to all its virtual legs; in contrast, applying a $Z$ operator to the physical leg corresponds to applying a $Z$ operator to any one of the virtual legs. The 3-leg tensor satisfies a $\mathbb{Z}_2$ parity constraint: its entries are 1 when the total parity of all indices is even, and 0 when the parity is odd. Applying an $X$ operator to the physical leg is equivalent to applying an $X$ operator to any one virtual leg; in contrast, applying a $Z$ operator to the physical leg corresponds to applying $Z$ operators to all virtual legs. (b) The tensor network operator for the usual duality transformation, which maps between the pure $\mathbb{Z}_2$ gauge theory satisfying $W_X^{[y]}=1$ and the Ising model with the periodic boundary condition. (c) The tensor network operator for the modified duality transformation , which maps between the pure $\mathbb{Z}_2$ gauge theory satisfying $W_X^{[y]}=-1$ and the Ising model with the twisted boundary condition.}
\label{Fig:app_duality}
\end{figure*}

\tableofcontents

\section{Roughening transition in view of the dual Ising model}\label{app:map_to_Ising}

For vanishing gauge-matter coupling $\lambda=0$, the $(2+1)$D $\mathbb{Z}_2$ gauge theory can be mapped to the $(2+1)$D quantum Ising model via a duality transformation~\cite{Wegner_duality_1971,Trebst_2007}. This duality relates a system of Ising spins (qubits) located on the lattice vertices to a system of $\mathbb{Z}_2$ gauge degrees of freedom (qubits) residing on the lattice edges. We denote the Hilbert space defined by qubits on the vertices (edges) of the square lattice as $\mathscr{H}_V$ ($\mathscr{H}_E$). The $\mathbb{Z}_2$ gauge theory at $\lambda=0$ is defined on $\mathscr{H}_E$ while the quantum Ising model is on $\mathscr{H}_V$. The duality transformation $\mathcal{D}$ can be written as a tensor network operator on the dual lattice~\cite{Duality_TN}, see Fig.~\ref{Fig:app_duality}b, and it satisfies: 
\begin{equation}\label{eq:duality_D}
    \mathcal{D}\mathcal{D}^{\dagger}=\prod_p \frac{1+A_v}{2} \frac{1+W^{[y]}_X}{2},
\end{equation}
so it is a non-invertible transformation. Because of the projectors in Eq.~\eqref{eq:duality_D}, when the system is on a torus, $\mathcal{D}$ maps the $\mathbb{Z}_2$ gauge theory at $\lambda=0$ without the electric charges and with only an even number of horizontal electric strings to the quantum Ising model. To map the $\mathbb{Z}_2$ gauge theory at $\lambda=0$ with an odd number of horizontal electric strings to the quantum Ising model, the duality transformation $\mathcal{D}$ needs to be modified by inserting $Z$ operators on the virtual level of the tensor network operator. We denote the modified duality transformation as $\mathcal{D}'$, see Fig.~\ref{Fig:app_duality}c, which satisfies
\begin{equation}
    \mathcal{D}'\mathcal{D}'^{\dagger}=\prod_p \frac{1+A_v}{2} \frac{1-W^{[y]}_X}{2}.
\end{equation}
Compared to Eq.~\eqref{eq:duality_D}, the sign in the 't Hooft loop projector is changed because the 't Hooft loop operator anti-commutes with the virtual $Z$ operators.
Using the symmetries of the tensors shown in Fig.~\ref{Fig:app_duality}a, it can be found that $\mathcal{D}'$ maps the $\mathbb{Z}_2$ gauge theory at zero-gauge matter coupling to the Ising model with a twisted boundary condition (vertical Ising spin couplings along a horizontal line change from ferromagnetic to anti-ferromagnetic):
\begin{equation}
    H_{\tI}=-h\sum_{\langle v v'\rangle}X_vX_{v'}+2h\sum_{\langle v(x,y_0) v'(x,y_0+1)\rangle}X_{v}X_{v'}-J_M\sum_v Z_v, 
\end{equation}
where $v(x,y)$ is a vertex whose coordinate is $(x,y)$. 

\begin{figure}
\centering
\includegraphics[width=0.99\linewidth]{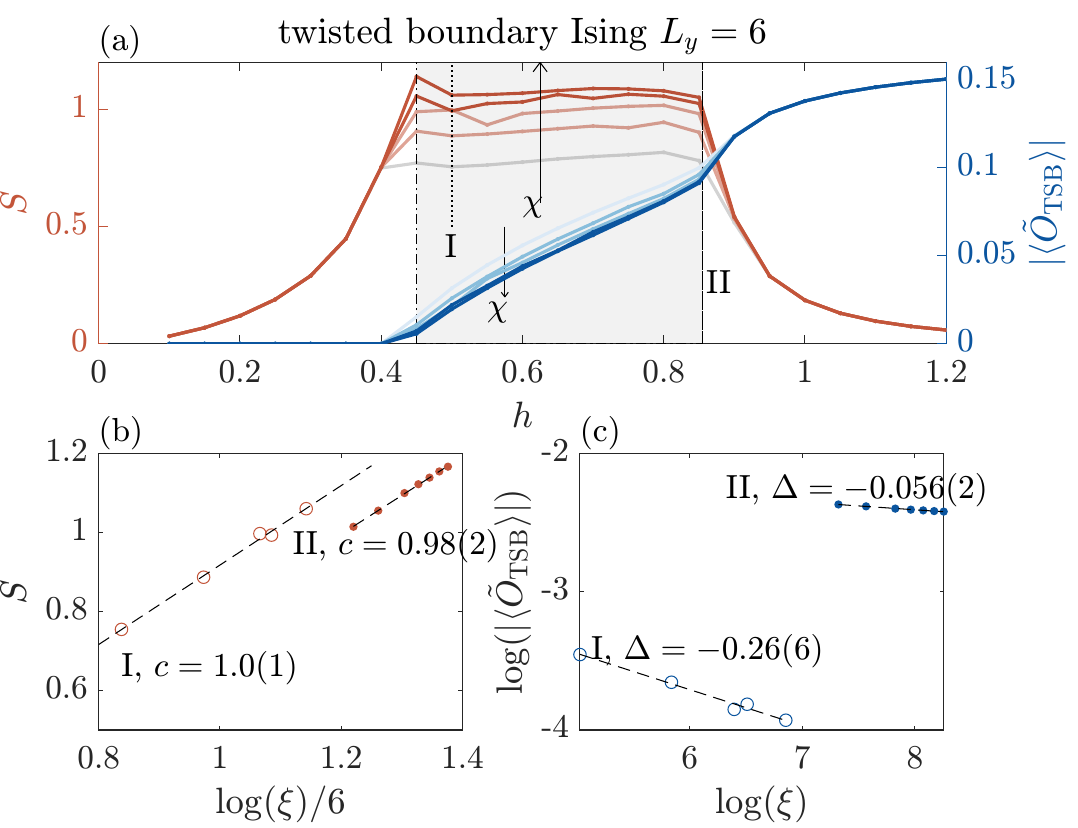}
\caption{\textbf{Twisted boundary Ising model.} (a) Entanglement entropy $S$ and the modified order parameter $|\langle O_{\tTSB}\rangle|$ obtained from iDMRG simulations with bond dimensions $\chi=100,200,\cdots,500$.  ``I'' is a cut at $h=0.5$, ``II'' is the estimated BKT transition point $h\approx0.855$. (b) Extracting the central $c$ from the entanglement entropy $S$ as a function of the correlation length $\xi$, where the bond dimenisons are $\chi=100,200,\cdots,500$ at the cut I and $\chi=400,500,\cdots,1000$ at the cut II. (c) Scaling of the order parameter $|\langle O_{\tTSB}\rangle|$ with the correlation length $\xi$, where $\chi=100,200,\cdots,500$ at the cut I and $\chi=400,500,\cdots,1000$ at the cut II.}
\label{Fig:app_Ising}
\end{figure}

We can also investigate the roughening transition by calculating the ground state of the twisted boundary Ising model using iDMRG. As shown in Fig.~\ref{Fig:app_Ising}a, for intermediate  $h$, the entanglement entropy of the states from the iDMRG simulations does not saturate with the bond dimension $\chi$, indicating a gapless phase. Fig.~\ref{Fig:app_Ising}b shows that the central charge extracted from the finite entanglement scaling is close to $1$, indicating that the gapless phase is a BKT phase. 

The twisted boundary Ising model has a modified translational symmetry in the direction $y$, which is generated by $\tilde{\mathcal{T}}_{\tI}^{[y]}=\mathcal{T}^{[y]}_{\tI}\prod_{x} X_{v(x,y_0)}$, 
where $\mathcal{T}^{[y]}_{\tI}$ is the usual translational operator along $y$ direction that maps site $v(x,y)$ to $v(x,y+1)$. 
Interestingly, $\mathcal{T}^{[y]}_{\tI}$ satisfies $(\mathcal{T}_{\tI}^{[y]})^{L_y}=1$ but $\tilde{\mathcal{T}}_{\tI}^{[y]}$ satisfies $(\tilde{\mathcal{T}}_{\tI}^{[y]})^{L_y}=\prod_v Z_v$. The order parameter detecting spontaneous symmetry breaking of $\tilde{\mathcal{T}}_{\tI}^{[y]}$ is $|\langle \tilde{O}_{\tTSB}\rangle|$, where
\begin{equation}
    \tilde{O}_{\tTSB}=\sum_y \exp\left( \frac{2\pi y}{L_y}\right)\frac{1-s_yX_{v(x,y)}X_{v(x,y+1)}}{2},
\end{equation}
with $s_y=1$ if $y\neq y_0$ and $s_y=-1$ if $y=y_0$. Actually, $\tilde{O}_{\tTSB}$ for the twisted boundary Ising model and $O_{\tTSB}$ for the $\mathbb{Z}_2$ gauge theory can be transformed with each other by the duality transformation $\mathcal{D}'$ defined in Fig.~\ref{Fig:app_duality}c. The order parameter $|\langle \tilde{O}_{\tTSB}\rangle|$ from the twisted boundary Ising model is shown in Fig.~\ref{Fig:app_Ising}a, which distinguishes the rough twisted boundary and the smooth twisted boundary. Moreover, at the expected BKT transition point, the scaling dimension of $\tilde{O}_{\tTSB}$ extracted from the iDMRG simulations is $\Delta\approx 0.0558$, see Fig.~\ref{Fig:app_Ising}c, consistent with the prediction $\Delta=2/L_y^2=1/18\approx 0.0556$ from the field theory. The results from the twisted boundary Ising model are consistent with those from the $\mathbb{Z}_2$ gauge theory in the main text. 

\section{Disorder parameter of the translational symmetry and Wilson loop correlator}

Usually, for a given global symmetry, we can define not only a local order parameter but also a non-local disorder parameter~\cite{Fradkin_2017_disorder_para}. For the twisted boundary Ising model, the disorder operator for the modified translational symmetry is simply the modified translational operator on a subsystem, and we denote it by $\tilde{\mathcal{T}}_{\tI}^{[y]}(x_1,x_2)$, where the subsystem is a finite cylinder whose two ends are at $x_1$ and $x_2$. 

In the paramagnetic phase of the twisted boundary Ising model, which corresponds to the deconfined phase of the $\mathbb{Z}_2$ gauge theory, $\langle \tilde{\mathcal{T}}_{\tI}^{[y]}(x_1,x_2)\rangle$ is finite when $|x_1-x_2|\rightarrow +\infty$ because $\tilde{\mathcal{T}}_{\tI}^{[y]}$ is not broken spontaneously. When the twisted boundary of the Ising model becomes smooth in the ferromagnetic phase, which corresponds to the stiff confined string of the $\mathbb{Z}_2$ gauge theory, $\langle\tilde{\mathcal{T}}_{\tI}^{[y]}(x_1,x_2)\rangle$ decays exponentially to zero with $|x_1-x_2|\rightarrow +\infty$ because $\tilde{\mathcal{T}}_{\tI}^{[y]}$ breaks spontaneously. When the twisted boundary of the Ising model becomes rough in the ferromagnetic phase, which corresponds to the floppy confined string of the $\mathbb{Z}_2$ gauge theory, $\tilde{\mathcal{T}}_{\tI}^{[y]}(x_1,x_2)$ algebraically decays to zero: $\langle\tilde{\mathcal{T}}_{\tI}^{[y]}(x_1,x_2)\rangle\sim |x_1-x_2|^{-\eta}$, where $\eta$ is the critical exponent of the correlation function.

However, for the $\mathbb{Z}_2$ lattice gauge theory, we cannot directly use the expectation value of the partial translation as a disorder parameter because the transition symmetry $\mathcal{T}^{[y]}$ fractionalizes when there is an electric string. This can be seen from the fact that the modified duality transformation $\mathcal{D}'$ in Fig.~\ref{Fig:app_duality}c maps $\mathcal{T}^{[y]}$ of the $\mathbb{Z}_2$ lattice gauge theory, which satisfies $\left(\mathcal{T}^{[y]}\right)^{L_y}=1$, to $\tilde{\mathcal{T}}_{\tI}^{[y]}$ of the twisted boundary Ising model, which satisfies $\left(\tilde{\mathcal{T}}_{\tI}^{[y]}\right)^{2L_y}=1$. We can find the correct disorder operator for the translational symmetry $\mathcal{T}^{[y]}$ of the $\mathbb{Z}_2$ lattice gauge theory using the disorder operator $\tilde{\mathcal{T}}_{\tI}^{[y]}(x_1,x_2)$ of the Ising model and the modified duality transformation $\mathcal{D}'$, which can be expressed as $\tilde{\mathcal{T}}^{[y]}(x_1,x_2)=\mathcal{D}'\tilde{\mathcal{T}}^{[y]}_{\tI}(x_1,x_2)\mathcal{D}'^{\dagger}$. 
And $\left\langle\tilde{\mathcal{T}}^{[y]}(x_1,x_2)\right\rangle$ is the disorder parameter for the translational symmetry of the $\mathbb{Z}_2$ lattice gauge theory.

Interestingly, $\tilde{\mathcal{T}}^{[y]}_{\tTC}(x_1,x_2)$ is related to two non-contractible Wilson loop operators: 
\begin{align}\label{Wilson_loop_corr}
    &\left[\tilde{\mathcal{T}}^{[y]}(x_1,x_2)\right]^{L_y}=\mathcal{D}'\left[\tilde{\mathcal{T}}^{[y]}_{\tI}(x_1,x_2)\right]^{L_y}\mathcal{D}'^{\dagger}\notag\\
    &=\mathcal{D}'\left[\prod_{\{v|v(x_1)\leq v(x)\leq v(x_2)\}}Z_v\right]\mathcal{D}'^{\dagger}=W^{[y]}_Z(x_1)W^{[y]}_Z(x_2),
\end{align}
 where $W_Z^{[y]}(x)= \prod_{e\in C^{[y]}(x)}Z_e$ is a non-contractible Wilson loop operator and $C^{[y]}(x)$ is vertical non-contractible loop on the primal lattice and labeled by the horizontal coordinate $x$. Consider the correlation function between two Wilson loop operators: $C_Z(x_1,x_2)=\left\langle W_Z^{[y]}(x_1)W_Z^{[y]}(x_2)\right\rangle$. In the deconfined phase, $\lim_{|x_1-x_2|\to \infty}C_Z\sim \exp(-\alpha L_y)$ with a constant $\alpha$ because of the perimeter law~\cite{Wegner_duality_1971,Kogut_LGH}, and $\lim_{|x_1-x_2|\to \infty}C_Z=0$ in the confined phase because of the area law. Importantly, we find the functional decay of the Wilson loop correlator $C_Z(x_1,x_2)$ can distinguish the strong and weak confinement at zero gauge-matter coupling. For strong confinement, $C_Z(x_1,x_2)$ decays exponentially with $|x_1-x_2|$, because the state with an electric string has a finite correlation length. For weak confinement, because the state with an electric string is in a BKT phase, the correlator decays algebraically: $C_Z(x_1,x_2) \sim|x_1-x_2|^{-\eta}$, where $\eta=2\Delta$ and $\Delta$ is the scaling dimension of $W^{[y]}_Z$. In the next section, we will derive the scaling scaling dimension of $W^{[y]}_Z$.   

\begin{table}
    \centering
    
    \begin{tabular}{C{2cm}C{1.5cm}C{1.5cm}C{1.7cm}C{1.5cm}}
    \toprule
          Latt. Operator &   $X_e-\langle X_e\rangle$ &  $O_{\text{TSB}}$ & $\left[\tilde{T}_{\tI}^{[y]}(x,+\infty)\right]^{N}$ & $W_Z^{[y]}$  \\  
           Field &   $\cos(L_y\sqrt{2}\theta)$ &  $\exp(i\sqrt{2}\theta)$ & $\exp(i\sqrt{2}\frac{N^2\phi}{L_y})$ & $\exp(i\sqrt{2}L_y\phi)$  \\ 
           \midrule
        $\Delta$ &  $L_y^2/R^2$  & $1/R^2$  & $N^2R^2/(4L_y^2)$ &  $R^2/4$   \\  
          $\Delta_{\text{BKT}}$ &  $2$  & $2/L_y^2$  & $N^2/8$&  $L_y^2/8$   \\ 
          \bottomrule
    \end{tabular}
    \caption{\textbf{Scaling dimensions.} Scaling dimensions $\Delta$ in the gapless BKT phase and $\Delta_{\text{BKT}}$ at the roughening transition, where $R$ is the Luttinger parameter, or equivalent the radius of a compactified free boson conformal field theory, and $L_y$ is the cylinder circumference. The electric field $\sum_e X_e$ drives the roughening transition, $O_{\tTSB}$ is the order parameter detecting the translational symmetry breaking in the strongly confined phase, and $W^{[y]}_Z$ is a non-contractible Wilson loop operator. $\left[\tilde{T}_{\tI}^{[y]}(x,+\infty)\right]^{N}$ is the disorder parameter of the translational symmetry of the twised boundary Ising model. Their corresponding fields are also shown.}
    \label{tab_app}
\end{table}

\section{Scaling dimensions of order parameters from the sine-Gordon theory}\label{app:SOP_and_ES}
\begin{figure}
\centering
\includegraphics[width=0.99\linewidth]{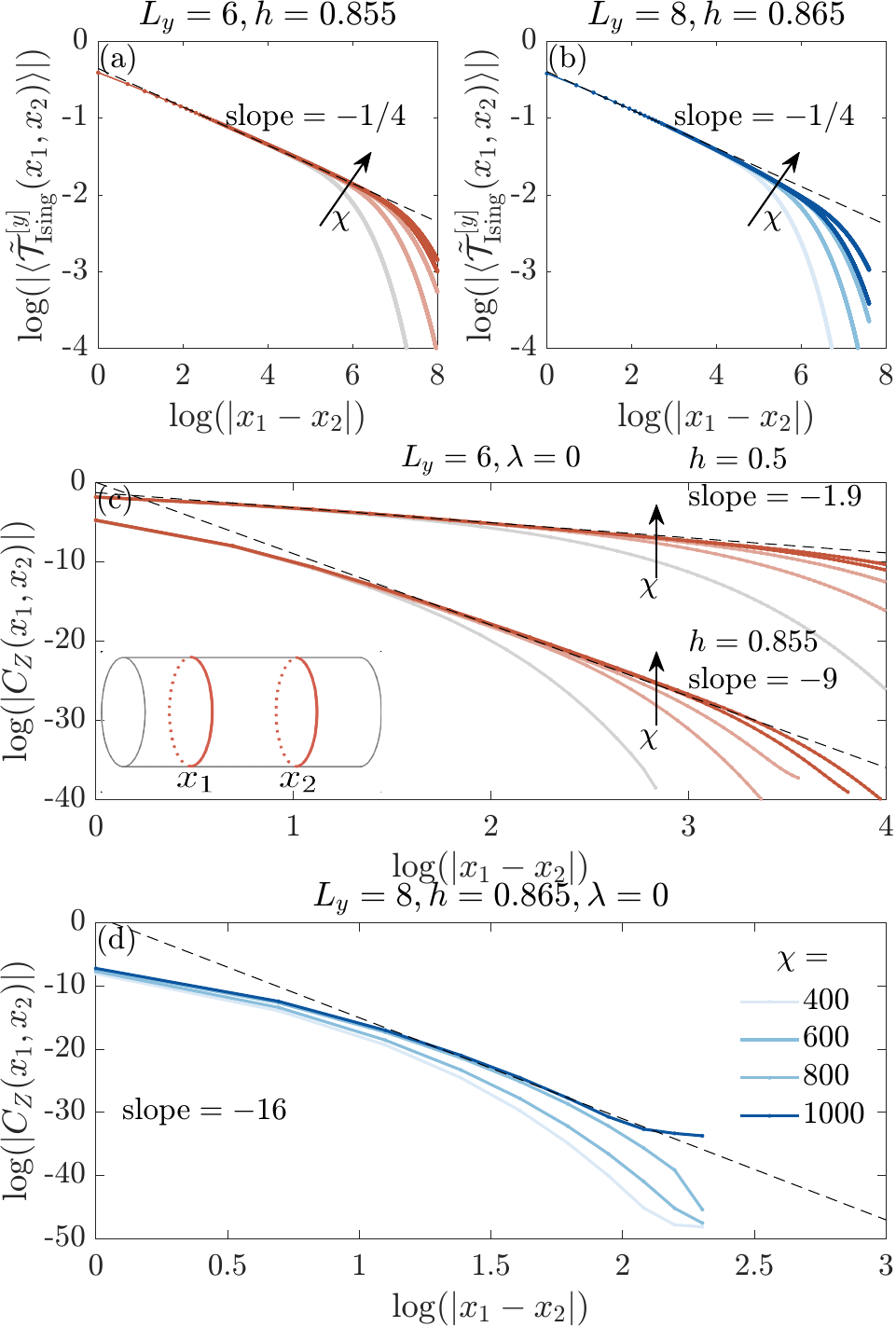}
\caption{\textbf{The disorder parameter of the twisted boundary Ising model and the Wilson loop correlator of the $\mathbb{Z}_2$ lattice gauge theory.} (a) Double-log plot of the disorder parameter $\langle \tilde{T}_{\tI}^{[y]}(x_1,x_2)\rangle$ for $L_y=6$, the results are obtained from iDMRG simulations with $\chi=200,400,\cdots,1000$. (b) The same as (a) but $L_y=8$. (c) Double-log plot of the correlation function between two non-contractible Wilson loop operators (see inset), calculated using iDMRG with $\chi=400,500,\cdots,1000$. The black dashed lines are predicted critical exponents $\eta$ from the theoretical scaling dimension of $W^{[y]}_Z$ in Tab.~\ref{tab_app}. Here, we set the cylinder circumference to $L_y=6$, and the magnetic excitation gap to $J_M=1$. (d) Double-log plot of the correlation function between two bare Wilson loop operators at the BKT transition point for $L_y=8$. The slope of the black dash line is the theoretically predicted exponent of the correlation function.}
\label{Fig:app_corr_non_local_op}
\end{figure}

Since the weakly confined regime belongs to the BKT phase and the $\mathbb{Z}_{L_y}$ translational symmetry is spontaneously broken in the strongly confined regime, the effective field theory describing the roughening transition of the confined string is the $\mathbb{Z}_{L_y}$ deformed sine-Gordon model~\cite{Wiegmann_1978,Matsuo_2006}:
\begin{align}\label{eq:deformed_SG}
    \mathcal{S}&=\frac{1}{\pi R^2} \int (\nabla\phi)^2 \, d^2 r+\frac{g_{\theta}}{2\pi \alpha^2} \int \cos({L_y\sqrt{2}\theta}) \, d^2 r,
\end{align}
where the real scalars $\phi$ and $\theta$, being compactified on a circle as $\phi=\phi+\sqrt{2}\pi$ (similarly for $\theta$), are mutually dual to each other, i.e., $\partial_x \phi = -R^2 \partial_y \theta/2$ and $\partial_y \phi = R^2 \partial_x \theta/2$, and $\alpha$ is an ultraviolet
cutoff, $g_{\theta}$ is a coupling constant and $R$ is the Luttinger parameter.
The last term in Eq.~\eqref{eq:deformed_SG} reduces the symmetry of the $\theta$ field from $U(1)$ [$\theta\rightarrow\theta+\gamma,\gamma\in [0,\sqrt{2}\pi)$] to $\mathbb{Z}_{L_y}$ ($\theta\rightarrow\theta+\sqrt{2}\pi m/L_y$ with $m=0,1,\cdots,L_y-1$). 

The field driving the BKT transition is $\cos(L_y\sqrt{2}\theta)$, whose scaling dimension is $\Delta\left[\cos(L_y\sqrt{2}\theta)\right]=L_y^2/R^2$. At the BKT transition, the field becomes marginally irrelevant so that $\Delta\left[\cos(L_y\sqrt{2}\theta)\right]=2$.  In the $\mathbb{Z}_2$ lattice gauge theory the electric field drives the roughening transition of the confined string. Thus,  we argue that the electric field corresponds the $\cos(L_y\sqrt{2}\theta)$ in the effective field theory.  Moreover, the order parameter of the $\mathbb{Z}_{L_y}$-deformed sine-Gordon model is $\exp(i\sqrt{2}\theta)$, because it transforms to $\exp(i\sqrt{2}\theta)\exp(i2\pi/L_y)$ by the $\mathbb{Z}_{L_y}$ symmetry transformation. Hence, we argue that $O_{\tTSB}$ of the $\mathbb{Z}_2$ lattice gauge theory 
correspond to the field of $\exp(i\sqrt{2}\theta)$ of the sine-Gordon model. The scaling dimension of $\exp(i\sqrt{2}\theta)$ is $\Delta\left[\exp(i\sqrt{2}\theta)\right]=1/R^2$, and at the BKT transition point, $\Delta\left[\exp(i\sqrt{2}\theta)\right]=2/L_y^2$ because $\Delta\left[\cos(L_y\sqrt{2}\theta)\right]=2$.  

In the deformed sine-Gordon model, $\cos(L_y\sqrt{2}\theta)$ and $\cos(\sqrt{2}\phi)$ are dual with each other~\cite{Matsuo_2006}, and the scaling dimension of $\cos(\sqrt{2}\phi)$ is $\Delta\left[\cos(\sqrt{2}\phi)\right]=R^2/4$. Similarly, from the perspective of the $\mathbb{Z}_2$ lattice gauge theory 
$O_{\tTSB}$ 
and the disorder operator $T^{[y]}(x_1,\infty)$ 
are ``dual'' with each other. So, the field in the effective field theory which corresponds to $T^{[y]}(x_1,\infty)$ ($\tilde{T}^{[y]}(x_1,\infty)$) is $\exp(i\sqrt{2}\phi/L_y)$, whose scaling dimension is $\Delta_{\exp(i\sqrt{2}\phi/L_y)}=R^2/(4 L_y^2)$. At the BKT transition point, $R^2=L_y^2/2$, so $\Delta\left[\exp(i\sqrt{2}\phi/L_y)\right]=1/8$, which is independent of $L_y$. Moreover, the exponent $\eta$ of the corelation function defined via $\langle\tilde{T}_{\tI}^{[y]}(x_1,x_2)\rangle\sim |x_1-x_2|^{-\eta'}$ can be derived: $\eta'=2\Delta\left[\exp(i\sqrt{2}\phi/L_y)\right]=1/4$. The same arguments hold for the dual Ising theory. In Fig.~\ref{Fig:app_corr_non_local_op}, we numerically calculate $\langle\tilde{\mathcal{T}}_{\tI}^{[y]}(x_1,x_2)\rangle$ using the iMPS with various bond dimensions to approximate the ground state of the twisted boundary of the Ising model, and we use the double-log plot to verify $\eta'=1/4$. 

Let us now consider the critical exponent of the correlation function between Wilson loops. From Eq.~\eqref{Wilson_loop_corr}, we consider the operator $\left[\tilde{\mathcal{T}}^{[y]}_{\tI}(x,+\infty)\right]^{N}$, which correspond to the field $\exp(i\sqrt{2}\phi N^2/L_y)$ with a scaling dimension $\Delta\left[\exp(i\sqrt{2}N^2\phi/L_y)\right]=N^2R^2/(4 L_y^2)$, and at the BKT transition point, $\Delta\left[\exp(i\sqrt{2}N^2\phi/L_y)\right]=N^2/8$. So $W^{[y]}_Z$ corresponds to $\exp(i\sqrt{2}\phi L_y)$ ($N=L_y$) with the scaling dimension $R^2/4$, and $\langle W_Z^{[y]}(x_1)W_Z^{[y]}(x_2)\rangle\sim |x_1-x_2|^{-R^2/2}$, where the exponent becomes $L^2/4$. In Figs.~\ref{Fig:app_corr_non_local_op}c and d, we show the Wilson loop correlator from the systems with the circumferences $L_y=6$ and $8$, from which we can find the critical exponent $\eta$ of the the Wilson loop correlator obtained from iDMRG simulations agrees with the theoretical predictions $L_y^2/4$ at the BKT transition points. Additionally, we summarize the operator-field correspondence and their scaling dimensions in Tab.~\ref{tab_app}.

We now study the limit of $L_y\rightarrow\infty$. To deal with the last term in Eq.~\eqref{eq:deformed_SG}, we rescale the field $\theta'=\sqrt{2}R\theta$ and $\phi'=\phi/(\sqrt{2}R)$ and use the relation  $\partial_x \phi = -R^2 \partial_y \theta/2$ to obtain:
\begin{align}
S=\frac{1}{8\pi} \int (\nabla\theta')^2 \, d^2 r+\frac{g_2}{2\pi \alpha^2} \int \cos(L_y\theta'/R) \, d^2 r,
\end{align}
where we have $\theta'=\theta'+2\pi R$ with the compactified radius $R$, since the coefficient of the first term is $1/(8\pi)$.
Near the BKT transition, $R^2/L_y^2$ is close to $1/2$, and the compactified radius $R$ goes to infinity with $L_y\rightarrow +\infty$.  
Thus the theory reduces to the free boson field theory without compactification (see Appendix 5.A. of Ref.~\cite{yellow_book_CFT}).

\section{Sequential iDMRG simulations}

\begin{figure}[!t]
\centering
\includegraphics[]{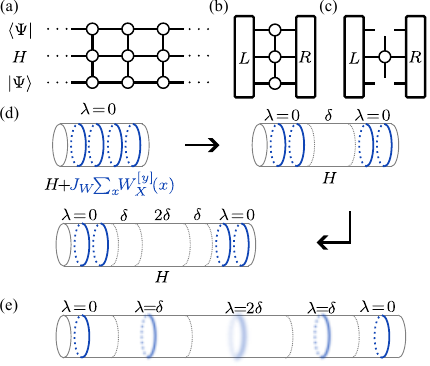}
\caption{\textbf{Sequential iDMRG simulations.} (a) In iDMRG, the Hamiltonian is expressed as an MPO and the ground state $\ket{\Psi}$ is an MPS. (b) Contracting the left and right parts of the tensor network in (a) gives rise to the left environment $L$ and the right environment $R$, separately. (c) The effective Hamiltonian, and the up (down) three open legs consist of the row (column) index of the effective Hamiltonian matrix. (d) The growth of the Hamiltonian MPO during all sweeps of the sequential iDMRG simulations, between the black dotted lines are different segments. A segment labeled by $n\delta$ is obtained from $(n+1)$-iDMRG simulation. (e) When pulling a 't Hooft loop through the entire MPO from one end to the middle, it changes from a bare 1-form symmetry to an emergent 1-form symmetry whose width is increasing.}
\label{Fig:app_explian_iDMRG}
\end{figure}

\begin{figure}[tbp]
\centering
\includegraphics[width=0.99\linewidth]{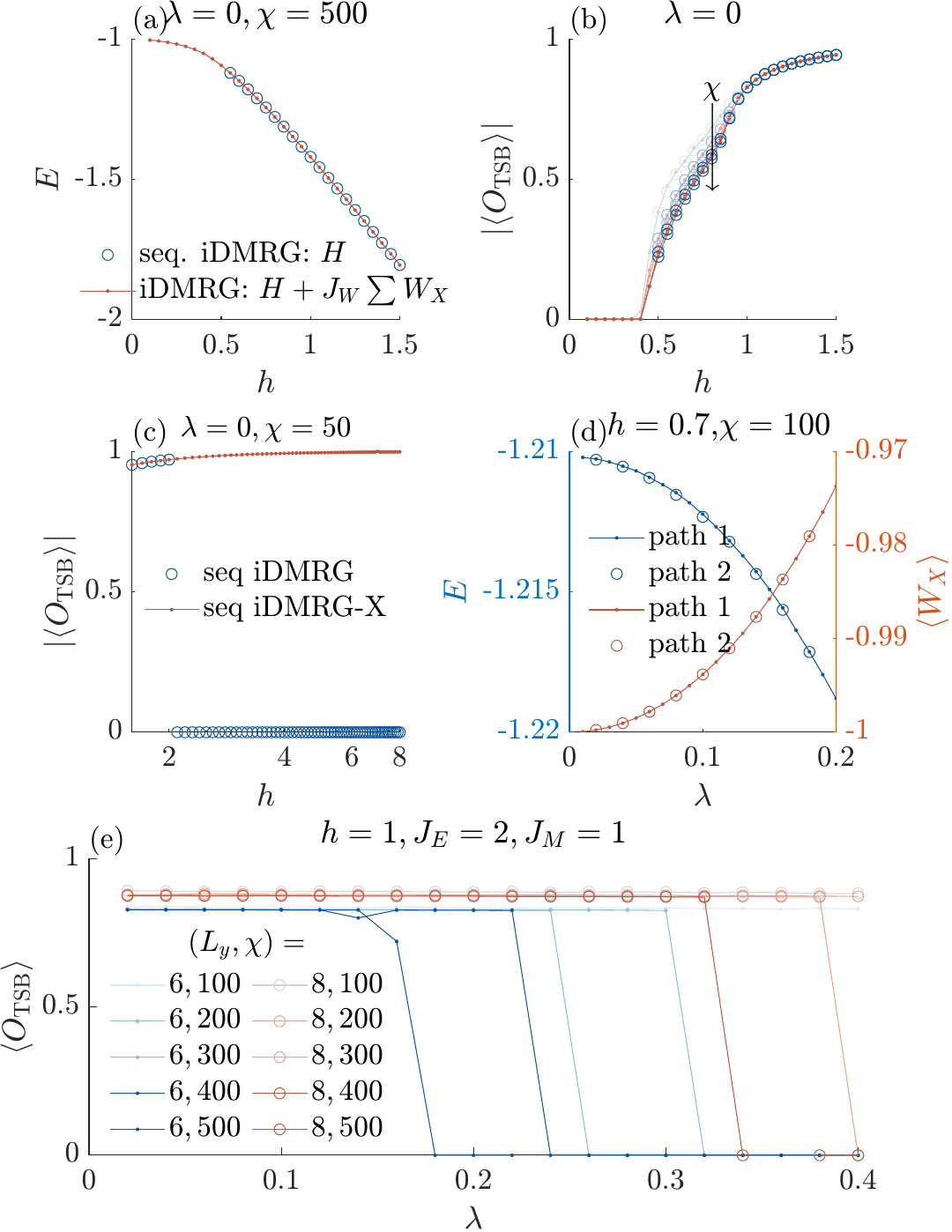}
\caption{\textbf{Validating the sequential iDMRG simulations.} (a) The energy density $E=\langle H\rangle/(L_xL_y)$ from iMPS with $\chi=500$ and $L_y=6$. We compare the results obtained by directly calculating the ground state of $H+J_W\sum_xW_X^{[y]}(x)$ and performing the sequential iDMRG simulations using $H$, separately, finding excellent agreement. (b) The translational symmetry breaking order parameter $|\langle O_{\tTSB}\rangle|$ from iMPS with $\chi=100,200,\cdots,500$ and $L_y=6$.  We compare the results obtained by directly calculating the ground state of $H+J_W\sum_xW_X^{[y]}(x)$ and performing the sequential iDMRG simulations using $H$, separately, finding excellent agreement. (c) Comparison of the 
translational symmetry breaking order parameter $|\langle O_{\tTSB}\rangle|$ obtained from the sequential iDMRG simulations with and without the assistance of DMRG-X for $L_y=6$. (d) Comparison of the ground state energy density $E$ and the expectation value of the bare 't Hooft loop operator $W_X$ obtained using sequential simulations along two different paths: 
Path 1: first calculate the ground state of $H(J_E=1,J_M=1, h,\lambda=0)+J_W\sum_xW_X^{[y]}(x)$ and then set $J_W=0$, and gradually increase $\lambda$; Path 2: first calculate the ground state of $H(J_E=1,J_M=1, h,\lambda>0+J_W\sum_xW_X^{[y]}(x)$, the gradually turn off $J_W$. (e) 
Translational symmetry breaking order parameter $|\langle O_{\tTSB}\rangle|$ obtained from sequential iDMRG with various $\chi$ and $L_y$.} 
\label{Fig:app_validate}
\end{figure}

In this section, we show the details of simulating confinement at finite gauge matter coupling ($\lambda\neq 0$) using sequential iDMRG. In iDMRG, the Hamiltonian is expressed in terms of a matrix product operator (MPO), and the ground state is expressed in terms of a matrix product state (MPS), as shown in Fig.~\ref{Fig:app_explian_iDMRG}a. During the iDMRG sweeps, the system size grows from the middle to the left and right side, such that the tensor network in Fig.~\ref{Fig:app_explian_iDMRG}a becomes longer and longer. When iDMRG converges, it returns a unit cell of tensors which can be used to build an infinite MPS (iMPS). The contraction of the left (right) part of the network gives rise to the environment $L$ $(R)$, see Fig.~\ref{Fig:app_explian_iDMRG}b. The effective Hamiltonian is shown in  Fig.~\ref{Fig:app_explian_iDMRG}c, whose ground state is a tensor in the unit cell of the iMPS.
For the details of the iDMRG algorithm, we refer to Ref.~\cite{tenpy}. The difficulty of simulating confinement at non-zero gauge-matter coupling arises from the fact that the 1-form 't Hooft loop symmetry is an emergent symmetry whose expression is unknown. As mentioned in the main text, this problem can be solved using the sequential iDMRG simulations, which mimics the adiabatic evolution of the eigenstate containing an electric string when tuning the gauge-matter coupling from $\lambda=0$ to $\lambda\neq 0$. 

In order to understand how the sequential iDMRG simulations work, we first consider the case $\lambda=0$ where the 1-form symmetry is exact and represented as a bare 't Hooft loop operator. We then calculate the ground state of $H(J_E,J_M,h,\lambda=0)+J_W\sum_x W_X^{[y]}(x)$, for sufficiently large $J_W$ and $J_E$ (the precise values do not matter due to the commuting structure). We then perform sufficiently many sweeps such that iDMRG converges. Following that, we initialize the second iDMRG simulation using the converged environment from the first iDMRG simulation and the Hamiltonian $H(J_E,J_M,h,\lambda=0)$ without the term $J_W\sum_x W_X^{[y]}(x)$. During the sweeps of the second iDMRG simulation, the iMPS always contains an electric string and never jumps to the ground state without the electric string, because of the topological properties of the 't Hooft loop ensuring $\langle W_X\rangle=-1$ from the initializing iDMRG simulation. In such a case, the entire MPO of two simulations is a Hamiltonian on a long cylinder with non-contractible 't Hooft loop operators at its two ends but not in the middle, as shown by the first two steps of Fig.~\ref{Fig:app_explian_iDMRG}d. 

.

Next, we consider the case that $\lambda=0$ but the electric field $h$ gradually increasing during the sequential simulation. We perform the first iDMRG simulation using the Hamiltonian $H(J_E,J_M,h,\lambda=0)+J_W\sum_x W_X^{[y]}(x)$ with the enough large $J_W$ and $J_E$, then perform the $n$-th iDMRG simulation using the Hamiltonian $H(J_E,J_M,h+(n-1)\delta,\lambda=0)$ without the term $J_W\sum_x W_X^{[y]}(x)$, where $n\geq 2$ and $\delta$ is small. We initialize the environment of the $n$-th iDMRG simulation using the converged environment of the $(n-1)$-th iDMRG simulation. 
In such sequential iDMRG simulations, the entire MPO is a Hamiltonian with different electric fields in different segments, see Fig.~\ref{Fig:app_explian_iDMRG}d, and the length of each segment is proportional to the number of the iDMRG sweeps. If $n$ is not too large, the $n$-th iDMRG simulation returns the iMPS at $h+\delta (n-1)$ with the electric string, i.e., $\langle W_X\rangle=-1$, which is just the ground state of $H(J_E,J_M,h+\delta(n-1),\lambda=0)+J_W\sum_x W_X^{[y]}(x)$ with a sufficiently large $J_W$, as shown in Figs.~\ref{Fig:app_validate}a and b.

However, when $n$ is too large, the electric field $h$ is very strong and the energy of the confined electric string is too high, so the sequential iDMRG could fail by returning an iMPS satisfying $\langle O_{\tTSB}\rangle=0$, see Fig.~\ref{Fig:app_validate}c, which implies $\langle W_X\rangle=1$. This problem can be circumvented by targeting the eigenstate of the effective Hamiltonian shown in Fig.~\ref{Fig:app_explian_iDMRG}c which has the largest overlap with the one from the last sweep, corresponding to DMRG-X used to calculate the highly excited states of systems with many-body localization~\cite{DMRG_X_2016,DMRG_X_2017}. Performing the sequential iDMRG-X simulations we find good convergence even for large $h$, Fig.~\ref{Fig:app_validate}c. 
When using DMRG-X, we have to use the single-site DMRG rather than the two-site DMRG, otherwise the targeting eigenstate of effective Hamiltonian can be degenerate and the simulations fail.

Equipped with this, we now consider finite gauge-matter coupling $\lambda\neq 0$, for which the 't Hooft loop symmetry is an emergent 1-form symmetry. We perform the first iDMRG simulation using $H(J_E,J_M,h,\lambda=0)+J_W\sum_x W_X^{[y]}(x)$ with sufficiently large $J_W$ and $J_E$, then perform the $n$-th iDMRG simulation using $H(J_E,J_M,h,\lambda=\delta (n-1))$ by initializing the environment of the $n$-th iDMRG simulation with the converged environment from the $(n-1)$-th iDMRG simulation, where $n\geq 2$. The DMRG-X method can be used to stabilize the sequential simulations. Similarly to the case $\lambda=0$,  the entire MPO is a Hamiltonian with different $\lambda$ in different segments, see Fig.~\ref{Fig:app_explian_iDMRG}d. 
Unlike the case $\lambda\neq 0$, when we pull the exact 1-form 't Hooft loop at the boundary segment to the bulk segments, see in Fig.~\ref{Fig:app_explian_iDMRG}e, it becomes the emergent 1-form 't Hooft loop symmetry, because 1-form symmetries are also robust against general perturbations. 
So, if $n$ is not too large, corresponding to $\lambda$ being not too large, the $n$-th sequential iDMRG simulation returns an iMPS with an electric string at $\lambda=\delta (n-1)$. However, when $\lambda$ is too large, the sequential iDMRG could fail by returning an iMPS without the electric string. This cannot be avoided even using the DMRG-X method. We observe that the failure of the simulation is caused by the approximate degeneracy of the targeting eigenstate. This is reasonable because the sequential iDMRG simulations mimic the adiabatic evolution, which becomes no longer valid when level crossings happen. Physically, we argue that this is related to the fact that there is no emergent 1-form 't Hooft loop symmetry when $\lambda$ is very large~\cite{Adam_Nahum_2021,Wen_emergent_high_form_2023,QEC_1_form_2025}. More specifically, for given $(J_E,J_M,h)$, there exist a threshold $\lambda_c$. When $\lambda<\lambda_c$, the system has the emergent 1-form 't Hooft loop symmetry, while the system does not have the emergent 1-form 't Hooft loop symmetry when $\lambda>\lambda_c$.  We expect that the sequential iDMRG simulations fail at some $\lambda<\lambda_c$, the main reason is that we consider a finite $L_y$ and we need to consider a large enough $L_y$ to get closer to $\lambda_c$. When $\lambda>\lambda_c$, \st{it} the emergent 't Hooft loop symmetry is absent, and the notion of confinement ceases to exist.

Some comments on the sequential simulations: First, we can in principle perform the sequential simulations along any path to the destination point in the parameter space, and the results obtained from different paths are the same as long as no phase transition is crossed along the paths and the starting points of different paths are in the same phase. For example, to obtain the eigenstate with the electric string at $\lambda\neq0$, we can also calculate the ground state of $H(J_E,J_M, h,\lambda>0)+J_W\sum_x W_X^{[y]}(x)$ and then slow turn off $J_W$. The result is the same as calculating the ground state of $H(J_E,J_M, h,\lambda=0)+J_W\sum_x W_X^{[y]}(x)$ first and then turn off $J_W$ and gradually turn on $\lambda$, see Fig.~\ref{Fig:app_validate}d. Moreover, when the bond dimension $\chi$ becomes larger, it becomes harder to reach larger $\lambda$, as shown in Fig.~\ref{Fig:app_validate}e, this reasonable because with a larger $\chi$, the gap of effective Hamiltonian becomes smaller, increasing the probability of the level crossing in the effective Hamiltonian. However, we find that when $L_y$ becomes larger it becomes easier to reach larger $\lambda$, see Fig.~\ref{Fig:app_validate}e. So, to reach a $\lambda$ (within the regime with emergent 1-form 't Hooft loop symmetry), we need to consider a balance between $\chi$ and $L_y$.

\section{$\mathbb{Z}_2$ lattice gauge theory on a cylinder with circumference $L_y=8$}
\begin{figure}[tbp]
\centering
\includegraphics[width=0.99\linewidth]{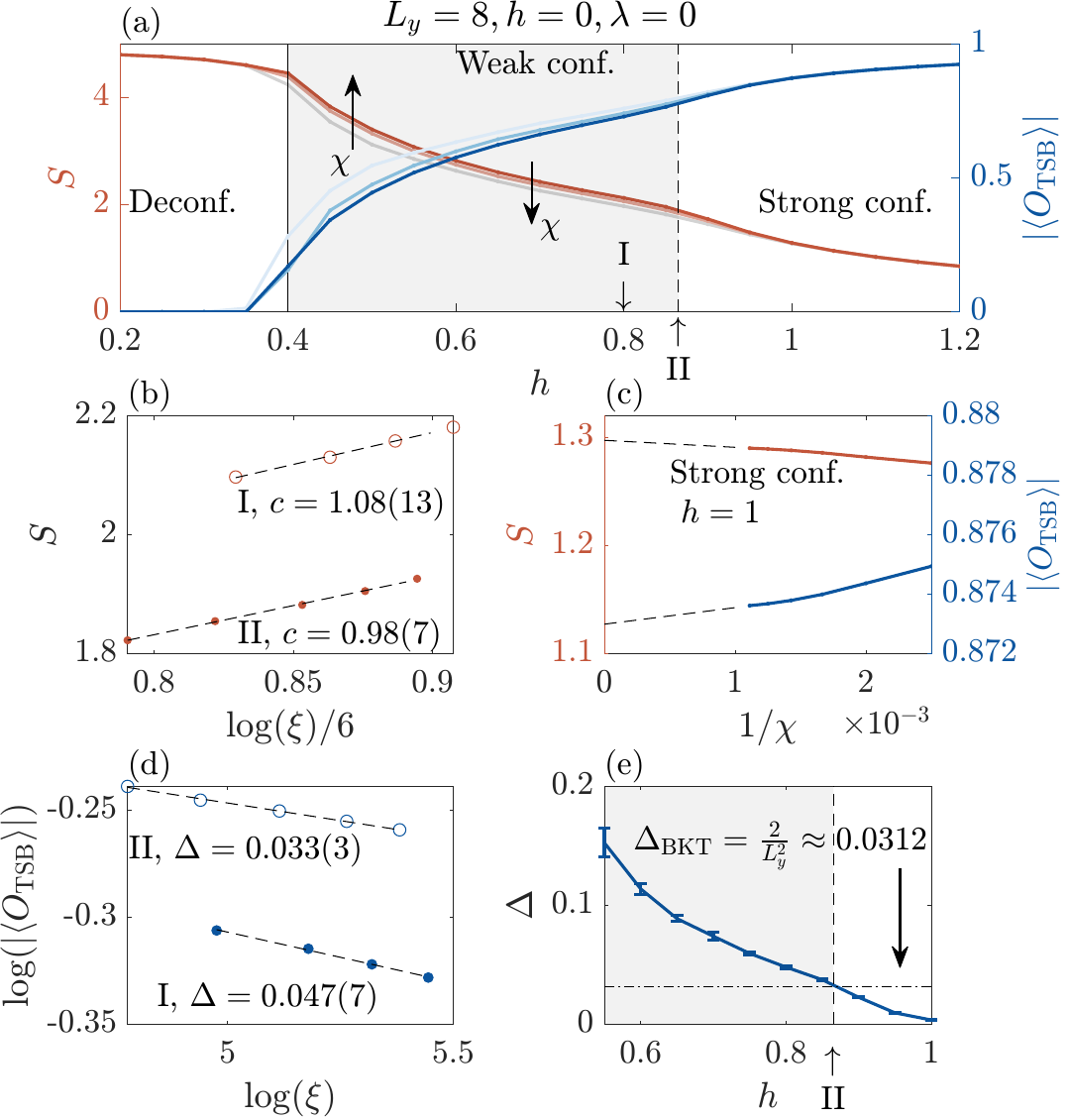}
\caption{\textbf{Roughening transition for zero gauge-matter coupling ($\lambda=0$) and $L_y=8$.} Here we set $J_M=1$ and perform the iDMRG simulations using the Hamiltonian $\tilde{H}$ and various bond dimensions $\chi$. (a) The entanglement entropy $S$ and the translational symmetry breaking order parameter $|\langle O_{\tTSB}\rangle|$ from the iDMRG simulations with $\chi=400,600,800$. ``I'' is a cut at $h=0.8$, ``II'' is the estimated BKT transition point $h\approx0.865$. (b) The central $c$ extracted from the iDMRG simulations with $\chi=700,800,900,1000$ at the cut I and $\chi=600,700,\cdots,1000$ at the cut II, $\xi$ is the correlation length induced by the finite bond dimensions. (c) $S$ and $|\langle O_{\tTSB}\rangle|$ as a function of $\chi$ at $h=1$. The dashed lines represent extrapolations to $\chi\rightarrow\infty$. (d) The order parameter $|\langle O_{\tTSB}\rangle|$ from the iDMRG simulations, where the bond dimensions are the same as those in (b). (e) The extracted scaling dimension $\Delta$ of $|\langle O_{\tTSB}\rangle|$ as a function of $h$. The dash-dot line indicates the scaling dimension $\Delta_{\text{BKT}}$ of $|\langle O_{\tTSB}\rangle|$ at the roughening (BKT) transition point.  
}
\label{Fig:app_hx_0_Ly_8}
\end{figure}
In the main text we show the results obtained from systems on a cylinder with a circumference $L_y=6$. In order to show how the BKT transition point changes with the circumference and that the relation between the scaling dimensions and $L_y$ is correct, we present the results obtained from systems on a cylinder with a circumference $L_y=8$ in this section.

We first consider the case of zero gauge-matter coupling, i.e., $\lambda=0$ where the 't Hooft loop symmetry is a bare 1-form symmetry. 
Fig.~\ref{Fig:app_hx_0_Ly_8}a shows the entanglement entropy $S$ and the translational symmetry breaking order parameter $|\langle O_{\tTSB}\rangle|$ from the iDMRG simulations with various bound dimensions. We find the existence of the floppy string in the middle where both $S$ and $|\langle O_{\tTSB}\rangle|$ do not saturate with the bond dimension $\chi$. In Fig.~\ref{Fig:app_hx_0_Ly_8}b, it can be found that the central charge $c$ extracted from the entanglement entropy $S$ and the correlation length $\xi$ is still close to $1$ when the confined string is floppy, indicating that the corresponding excited state still belongs to the BKT phase. Compared to the central charge extracted at $L_y = 6$ with $\chi \leq 1000$, the extracted central charge at $L_y = 8$ and  $\chi \leq 1000$ has a larger error. This is because, in the weakly confined phase, the entanglement entropy follows the form $S = \alpha L_y + \frac{c}{6} \log \xi+\cdots$ only when the bond dimension is sufficiently large. If $\chi$ is too small, the first area law term of the entanglement entropy does not saturate, leading to a larger error in the central charge extracted from the second finite correlation scaling term. In Fig.~\ref{Fig:app_hx_0_Ly_8}c, at $h = 1$, both the entanglement entropy and the order parameter saturate to finite values as $\chi \rightarrow \infty$, indicating that the electric flux tube is strongly confined. In Figs.~\ref{Fig:app_hx_0_Ly_8}d and e, we extract the scaling dimension of $O_{\tTSB}$. From the scaling dimension $\Delta=2/L_y^2$ at the BKT transition point, we determine that the roughening transition point is at $h\approx0.865$, where the theoretical scaling dimension is $2/64\approx 0.0312$, which matches the numerical result $0.033(3)$. The discrepancy between the numerically extracted scaling dimension and the theoretical value arises from not only the limited bond dimension, but also the uncertainty in the location of the BKT transition point. The location $h\approx0.865$ of the BKT transition at $L_y=8$ is slightly larger than  $h\approx0.855$ at $L_y=6$ and zero gauge-matter coupling.

\begin{figure}[tbp]
\centering
\includegraphics[width=0.99\linewidth]{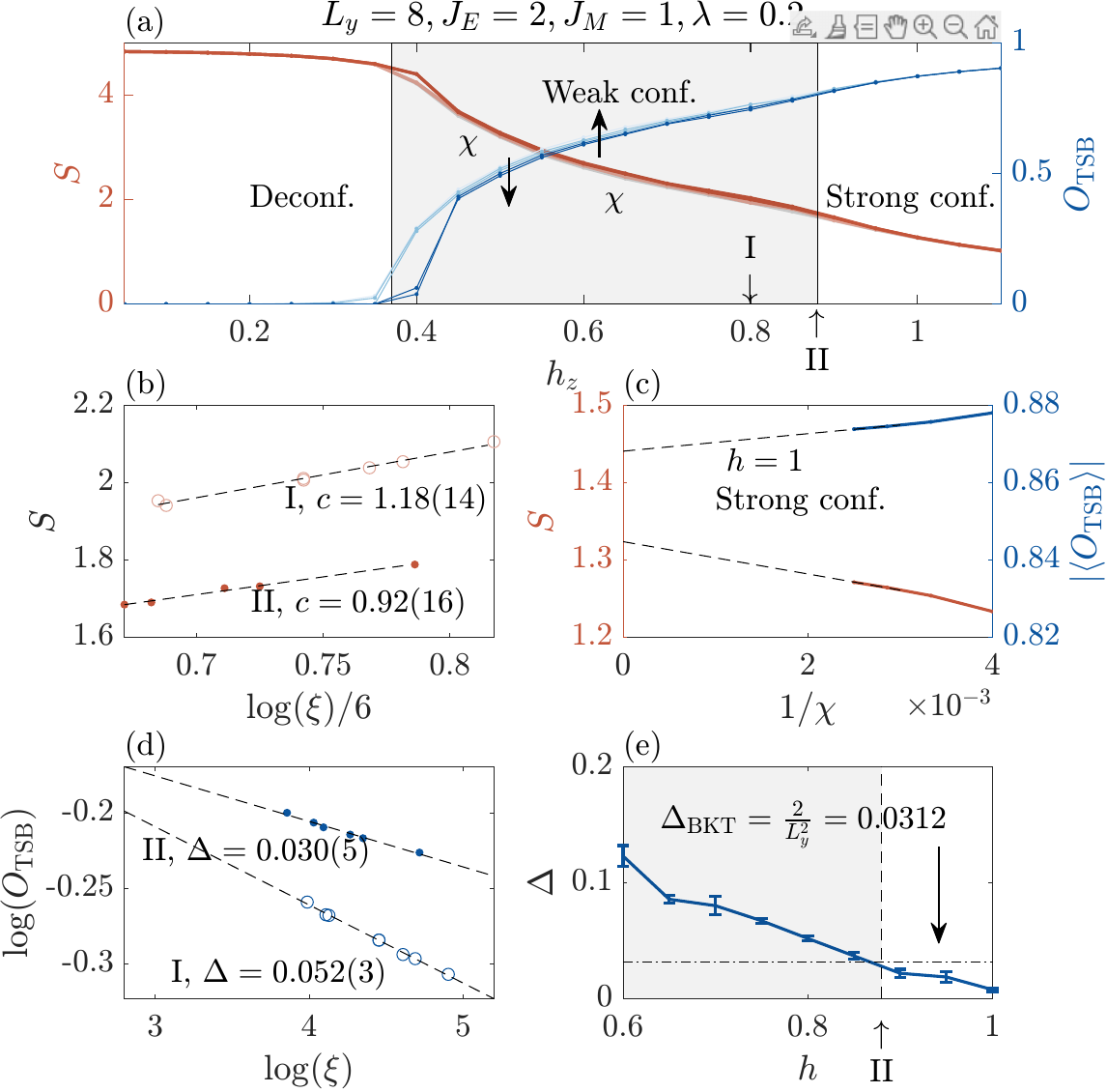}
\caption{\textbf{Roughening transition for finite gauge-matter coupling and $L_y=8$.} Here we set $\lambda=0.2$ and $J_M=1$ and $J_E=2$ and perform the sequential iDMRG simulations with various bond dimensions $\chi$. (a) The entanglement entropy $S$ and the translational symmetry breaking order parameter $|\langle O_{\tTSB}\rangle|$ from the sequential iDMRG simulations with $\chi=350,400,\cdots,550$. ``I'' is a cut at $h=0.8$, ``II'' is the estimated BKT transition point $h\approx0.88$. (b) The central $c$ extracted from the sequential iDMRG simulations with $\chi=350,400,\cdots,650$ at the cut I and $\chi=350,400,\cdots,550$ at the cut II, $\xi$ is the correlation length induced by the finite bond dimensions. (c) $S$ and $|\langle O_{\tTSB}\rangle|$ as a function of $\chi$ at $h=1$. The dashed lines represent extrapolations to $\chi\rightarrow\infty$. (d) The order parameter $|\langle O_{\tTSB}\rangle|$ from the iDMRG simulations, whose bond dimensions are the same as those in (b). (e) The extracted scaling dimension $\Delta$ of $|\langle O_{\tTSB}\rangle|$ as a function of $h$. The dash-dot line indicates the scaling dimension $\Delta_{\text{BKT}}$ of $|\langle O_{\tTSB}\rangle|$ at the roughening (BKT) transition point. } 
\label{Fig:app_hx_02_Ly_8}
\end{figure}

Next let us consider finite gauge-matter coupling $\lambda=0.2$, and $J_E=2$ and $J_M=1$ and $L_y=8$. Fig.~\ref{Fig:app_hx_02_Ly_8}a shows the entanglement entropy $S$ and the translational symmetry breaking order parameter $|\langle O_{\tTSB}\rangle|$ from the sequential iDMRG simulations with various bound dimensions. We find the existence of the floppy string in the middle where both $S$ and $|\langle O_{\tTSB}\rangle|$ do not saturate with the bond dimensions $\chi$. In Fig.~\ref{Fig:app_hx_02_Ly_8}b, it can be found that the central charge $c$ extracted from the entanglement entropy $S$ and the correlation length $\xi$ is still close to $1$ when the confined string is floppy, indicating that the corresponding excited state still belongs to a BKT phase. In Fig.~\ref{Fig:app_hx_02_Ly_8}c, at $h = 1$, both the entanglement entropy and the order parameter saturate to finite values as $\chi \rightarrow \infty$, indicating that the system is in the strongly confined regime. In Figs.~\ref{Fig:app_hx_02_Ly_8}d and e, we extract the scaling dimension of $O_{\tTSB}$. Since the scaling dimension is $\Delta=2/L_y^2$ at the BKT transition point, we obtain the roughening transition point at $h\approx0.88$, where the theoretical scaling dimension is $2/64\approx 0.0312$ matching the respective numerical result $0.030(4)$. The location $h\approx0.88$ of the BKT transition at $L_y=8$ is only slightly larger than  $h\approx0.875$ obtained at $L_y=6,\lambda=0.2,J_E=2$ and $J_M=1$. 

Finally, we comment on the location of the phase boundary between the deconfined phase and the confined region. This phase boundary is described by the (2+1)D Ising conformal field theory~\cite{Adam_Nahum_2021}, which leads to strong finite circumference effects in iDMRG simulations. Since $\langle|O_{\tTSB}|\rangle$ vanishes in both the deconfined and weakly confined phases as $\chi\rightarrow\infty$, it cannot be used to detect this transition. Instead, for a given $L_y$, we locate the phase boundary by the peak of the entanglement entropy. At $\lambda = 0$, the peak occurs at $h \approx 0.45$ for $L_y = 6$ and $h \approx 0.4$ for $L_y = 8$, gradually approaching the theoretical value $h_c = 0.328474(3)$ as $L_y \rightarrow \infty$~\cite{Deng_2002}.

\section{$\mathbb{Z}_3$ clock model}
\begin{figure}[tbp]
\centering
\includegraphics[width=0.99\linewidth]{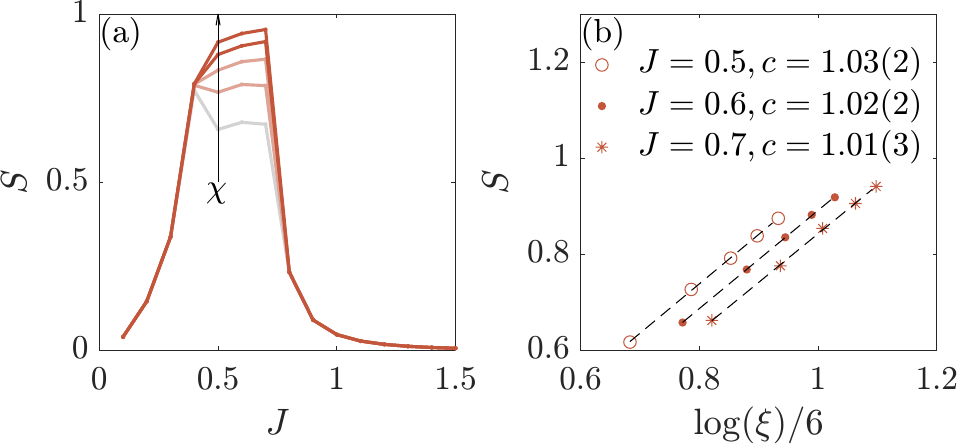}
\caption{\textbf{$\mathbb{Z}_3$ clock model} (a) The entanglement entropy from the iDMRG simulations with $\chi=100,200,\cdots,500$ approximating the ground state of the twisted boundary $\mathbb{Z}_3$ clock model shown in Eq.~\eqref{eq:Z3_clock_ham}. (b) Extracting the central charge $c$ from the finite entanglement scaling, where the bond dimensions for the iDMRG simulations are $100,200,\cdots, 500$ for different $J$ in the gapless phase.}
\label{Fig:app_Z_3}
\end{figure}

The roughening transition of confined strings exists in various gauge theories. Our method can be applied not only to  $\mathbb{Z}_2$ lattice gauge theories but also the other models. As an example, we consider the twisted boundary 2D $\mathbb{Z}_3$ quantum clock model:
\begin{align}\label{eq:Z3_clock_ham}
    H_{\text{clock}}&=-J\sum_{\langle v v'\rangle}\left(\tilde{X}_v\tilde{X}^{\dagger}_{v'}+\text{h.c.}\right)-\sum_v \left(\tilde{Z}_v+\text{h.c.}\right)\notag\\
    &+\sum_{\langle v(x,y_0) v'(x,y_0+1)\rangle}\left[(J-\omega J)\tilde{X}_{v}\tilde{X}^{\dagger}_{v'}+\text{h.c.}\right], 
\end{align}
where $\tilde{Z}=\sum_{n=0}^{2}\omega^n\ket{n}\bra{n}$ with $\omega=\exp(2i \pi n/3)$ and $\tilde{X}=\sum_{n=0}^{2}\ket{n}\bra{(n+1)\mod{3}}$ are the $\mathbb{Z}_3$ generalization of the $X$ and $Z$ operators, and $\{\ket{n}|n=0,1,2\}$ is the basis of a qudit. A $\mathbb{Z}_3$ generalization of the duality transformation shown in Fig.~\ref{Fig:app_duality}c can be defined, such that the $\mathbb{Z}_3$ clock model with the twisted boundary condition can be transformed to the $\mathbb{Z}_3$ lattice gauge theory with the electric strings. For simplicity, we consider the twisted boundary $\mathbb{Z}_3$ clock model on a cylinder and calculate the ground state using iDMRG. As shown in Fig.~\ref{Fig:app_Z_3}a, for intermediate $J$ it can be found that the entanglement entropy from iDMRG simulations does not saturate with $\chi$ increasing, similarly to the results of the twisted boundary Ising model, which  implies a gapless phase. In Fig.~\ref{Fig:app_Z_3}b, we use the finite entanglement scaling to extract the central charge in the gapless phase, which we find to be very close to one, indicating the gapless phase is a BKT phase. These results confirm that for a large class of lattice gauge theories, the weakly confined regimes belong to the BKT phase.  

%% file: main_revised_arXiv.bbl
\begin{thebibliography}{85}%
\makeatletter
\providecommand \@ifxundefined [1]{%
 \@ifx{#1\undefined}
}%
\providecommand \@ifnum [1]{%
 \ifnum #1\expandafter \@firstoftwo
 \else \expandafter \@secondoftwo
 \fi
}%
\providecommand \@ifx [1]{%
 \ifx #1\expandafter \@firstoftwo
 \else \expandafter \@secondoftwo
 \fi
}%
\providecommand \natexlab [1]{#1}%
\providecommand \enquote  [1]{``#1''}%
\providecommand \bibnamefont  [1]{#1}%
\providecommand \bibfnamefont [1]{#1}%
\providecommand \citenamefont [1]{#1}%
\providecommand \href@noop [0]{\@secondoftwo}%
\providecommand \href [0]{\begingroup \@sanitize@url \@href}%
\providecommand \@href[1]{\@@startlink{#1}\@@href}%
\providecommand \@@href[1]{\endgroup#1\@@endlink}%
\providecommand \@sanitize@url [0]{\catcode `\\12\catcode `\$12\catcode `\&12\catcode `\#12\catcode `\^12\catcode `\_12\catcode `\%12\relax}%
\providecommand \@@startlink[1]{}%
\providecommand \@@endlink[0]{}%
\providecommand \url  [0]{\begingroup\@sanitize@url \@url }%
\providecommand \@url [1]{\endgroup\@href {#1}{\urlprefix }}%
\providecommand \urlprefix  [0]{URL }%
\providecommand \Eprint [0]{\href }%
\providecommand \doibase [0]{https://doi.org/}%
\providecommand \selectlanguage [0]{\@gobble}%
\providecommand \bibinfo  [0]{\@secondoftwo}%
\providecommand \bibfield  [0]{\@secondoftwo}%
\providecommand \translation [1]{[#1]}%
\providecommand \BibitemOpen [0]{}%
\providecommand \bibitemStop [0]{}%
\providecommand \bibitemNoStop [0]{.\EOS\space}%
\providecommand \EOS [0]{\spacefactor3000\relax}%
\providecommand \BibitemShut  [1]{\csname bibitem#1\endcsname}%
\let\auto@bib@innerbib\@empty
\bibitem [{\citenamefont {Wilson}(1974)}]{Wilson_confinemnt_1974}%
  \BibitemOpen
  \bibfield  {author} {\bibinfo {author} {\bibfnamefont {K.~G.}\ \bibnamefont {Wilson}},\ }\bibfield  {title} {\bibinfo {title} {Confinement of quarks},\ }\href {https://doi.org/10.1103/PhysRevD.10.2445} {\bibfield  {journal} {\bibinfo  {journal} {Phys. Rev. D}\ }\textbf {\bibinfo {volume} {10}},\ \bibinfo {pages} {2445} (\bibinfo {year} {1974})}\BibitemShut {NoStop}%
\bibitem [{\citenamefont {McCoy}\ and\ \citenamefont {Wu}(1978)}]{McCoy1978}%
  \BibitemOpen
  \bibfield  {author} {\bibinfo {author} {\bibfnamefont {B.~M.}\ \bibnamefont {McCoy}}\ and\ \bibinfo {author} {\bibfnamefont {T.~T.}\ \bibnamefont {Wu}},\ }\bibfield  {title} {\bibinfo {title} {Two-dimensional ising field theory in a magnetic field: Breakup of the cut in the two-point function},\ }\href {https://doi.org/10.1103/physrevd.18.1259} {\bibfield  {journal} {\bibinfo  {journal} {Phys. Rev. D}\ }\textbf {\bibinfo {volume} {18}},\ \bibinfo {pages} {1259–1267} (\bibinfo {year} {1978})}\BibitemShut {NoStop}%
\bibitem [{\citenamefont {Rutkevich}(2008)}]{Rutkevich2008}%
  \BibitemOpen
  \bibfield  {author} {\bibinfo {author} {\bibfnamefont {S.~B.}\ \bibnamefont {Rutkevich}},\ }\bibfield  {title} {\bibinfo {title} {Energy spectrum of bound-spinons in the quantum ising spin-chain ferromagnet},\ }\href {https://doi.org/10.1007/s10955-008-9495-1} {\bibfield  {journal} {\bibinfo  {journal} {Journal of Statistical Physics}\ }\textbf {\bibinfo {volume} {131}},\ \bibinfo {pages} {917–939} (\bibinfo {year} {2008})}\BibitemShut {NoStop}%
\bibitem [{\citenamefont {Kormos}\ \emph {et~al.}(2017)\citenamefont {Kormos}, \citenamefont {Collura}, \citenamefont {Takács},\ and\ \citenamefont {Calabrese}}]{confinement_quench_2017}%
  \BibitemOpen
  \bibfield  {author} {\bibinfo {author} {\bibfnamefont {M.}~\bibnamefont {Kormos}}, \bibinfo {author} {\bibfnamefont {M.}~\bibnamefont {Collura}}, \bibinfo {author} {\bibfnamefont {G.}~\bibnamefont {Takács}},\ and\ \bibinfo {author} {\bibfnamefont {P.}~\bibnamefont {Calabrese}},\ }\bibfield  {title} {\bibinfo {title} {Real-time confinement following a quantum quench to a non-integrable model},\ }\href {https://doi.org/10.1038/nphys3934} {\bibfield  {journal} {\bibinfo  {journal} {Nature Physics}\ }\textbf {\bibinfo {volume} {13}},\ \bibinfo {pages} {246} (\bibinfo {year} {2017})}\BibitemShut {NoStop}%
\bibitem [{\citenamefont {Mazza}\ \emph {et~al.}(2019)\citenamefont {Mazza}, \citenamefont {Perfetto}, \citenamefont {Lerose}, \citenamefont {Collura},\ and\ \citenamefont {Gambassi}}]{Mazza2019}%
  \BibitemOpen
  \bibfield  {author} {\bibinfo {author} {\bibfnamefont {P.~P.}\ \bibnamefont {Mazza}}, \bibinfo {author} {\bibfnamefont {G.}~\bibnamefont {Perfetto}}, \bibinfo {author} {\bibfnamefont {A.}~\bibnamefont {Lerose}}, \bibinfo {author} {\bibfnamefont {M.}~\bibnamefont {Collura}},\ and\ \bibinfo {author} {\bibfnamefont {A.}~\bibnamefont {Gambassi}},\ }\bibfield  {title} {\bibinfo {title} {Suppression of transport in nondisordered quantum spin chains due to confined excitations},\ }\href {https://doi.org/10.1103/physrevb.99.180302} {\bibfield  {journal} {\bibinfo  {journal} {Phys. Rev. B}\ }\textbf {\bibinfo {volume} {99}},\ \bibinfo {pages} {180302} (\bibinfo {year} {2019})}\BibitemShut {NoStop}%
\bibitem [{\citenamefont {Liu}\ \emph {et~al.}(2019)\citenamefont {Liu}, \citenamefont {Lundgren}, \citenamefont {Titum}, \citenamefont {Pagano}, \citenamefont {Zhang}, \citenamefont {Monroe},\ and\ \citenamefont {Gorshkov}}]{Liu2019}%
  \BibitemOpen
  \bibfield  {author} {\bibinfo {author} {\bibfnamefont {F.}~\bibnamefont {Liu}}, \bibinfo {author} {\bibfnamefont {R.}~\bibnamefont {Lundgren}}, \bibinfo {author} {\bibfnamefont {P.}~\bibnamefont {Titum}}, \bibinfo {author} {\bibfnamefont {G.}~\bibnamefont {Pagano}}, \bibinfo {author} {\bibfnamefont {J.}~\bibnamefont {Zhang}}, \bibinfo {author} {\bibfnamefont {C.}~\bibnamefont {Monroe}},\ and\ \bibinfo {author} {\bibfnamefont {A.~V.}\ \bibnamefont {Gorshkov}},\ }\bibfield  {title} {\bibinfo {title} {Confined quasiparticle dynamics in long-range interacting quantum spin chains},\ }\href {https://doi.org/10.1103/PhysRevLett.122.150601} {\bibfield  {journal} {\bibinfo  {journal} {Phys. Rev. Lett.}\ }\textbf {\bibinfo {volume} {122}},\ \bibinfo {pages} {150601} (\bibinfo {year} {2019})}\BibitemShut {NoStop}%
\bibitem [{\citenamefont {Lerose}\ \emph {et~al.}(2020)\citenamefont {Lerose}, \citenamefont {Surace}, \citenamefont {Mazza}, \citenamefont {Perfetto}, \citenamefont {Collura},\ and\ \citenamefont {Gambassi}}]{Lerose2020}%
  \BibitemOpen
  \bibfield  {author} {\bibinfo {author} {\bibfnamefont {A.}~\bibnamefont {Lerose}}, \bibinfo {author} {\bibfnamefont {F.~M.}\ \bibnamefont {Surace}}, \bibinfo {author} {\bibfnamefont {P.~P.}\ \bibnamefont {Mazza}}, \bibinfo {author} {\bibfnamefont {G.}~\bibnamefont {Perfetto}}, \bibinfo {author} {\bibfnamefont {M.}~\bibnamefont {Collura}},\ and\ \bibinfo {author} {\bibfnamefont {A.}~\bibnamefont {Gambassi}},\ }\bibfield  {title} {\bibinfo {title} {Quasilocalized dynamics from confinement of quantum excitations},\ }\href {https://doi.org/10.1103/physrevb.102.041118} {\bibfield  {journal} {\bibinfo  {journal} {Phys. Rev. B}\ }\textbf {\bibinfo {volume} {102}},\ \bibinfo {pages} {041118} (\bibinfo {year} {2020})}\BibitemShut {NoStop}%
\bibitem [{\citenamefont {Verdel}\ \emph {et~al.}(2023)\citenamefont {Verdel}, \citenamefont {Zhu},\ and\ \citenamefont {Heyl}}]{zhu_2023}%
  \BibitemOpen
  \bibfield  {author} {\bibinfo {author} {\bibfnamefont {R.}~\bibnamefont {Verdel}}, \bibinfo {author} {\bibfnamefont {G.-Y.}\ \bibnamefont {Zhu}},\ and\ \bibinfo {author} {\bibfnamefont {M.}~\bibnamefont {Heyl}},\ }\bibfield  {title} {\bibinfo {title} {Dynamical localization transition of string breaking in quantum spin chains},\ }\href {https://doi.org/10.1103/PhysRevLett.131.230402} {\bibfield  {journal} {\bibinfo  {journal} {Phys. Rev. Lett.}\ }\textbf {\bibinfo {volume} {131}},\ \bibinfo {pages} {230402} (\bibinfo {year} {2023})}\BibitemShut {NoStop}%
\bibitem [{\citenamefont {Birnkammer}\ \emph {et~al.}(2022)\citenamefont {Birnkammer}, \citenamefont {Bastianello},\ and\ \citenamefont {Knap}}]{Prethermal_2022}%
  \BibitemOpen
  \bibfield  {author} {\bibinfo {author} {\bibfnamefont {S.}~\bibnamefont {Birnkammer}}, \bibinfo {author} {\bibfnamefont {A.}~\bibnamefont {Bastianello}},\ and\ \bibinfo {author} {\bibfnamefont {M.}~\bibnamefont {Knap}},\ }\bibfield  {title} {\bibinfo {title} {Prethermalization in one-dimensional quantum many-body systems with confinement},\ }\href {https://doi.org/10.1038/s41467-022-35301-6} {\bibfield  {journal} {\bibinfo  {journal} {Nature Communications}\ }\textbf {\bibinfo {volume} {13}},\ \bibinfo {pages} {7663} (\bibinfo {year} {2022})}\BibitemShut {NoStop}%
\bibitem [{\citenamefont {Simon}\ \emph {et~al.}(2011)\citenamefont {Simon}, \citenamefont {Bakr}, \citenamefont {Ma}, \citenamefont {Tai}, \citenamefont {Preiss},\ and\ \citenamefont {Greiner}}]{Simon2011}%
  \BibitemOpen
  \bibfield  {author} {\bibinfo {author} {\bibfnamefont {J.}~\bibnamefont {Simon}}, \bibinfo {author} {\bibfnamefont {W.~S.}\ \bibnamefont {Bakr}}, \bibinfo {author} {\bibfnamefont {R.}~\bibnamefont {Ma}}, \bibinfo {author} {\bibfnamefont {M.~E.}\ \bibnamefont {Tai}}, \bibinfo {author} {\bibfnamefont {P.~M.}\ \bibnamefont {Preiss}},\ and\ \bibinfo {author} {\bibfnamefont {M.}~\bibnamefont {Greiner}},\ }\bibfield  {title} {\bibinfo {title} {Quantum simulation of antiferromagnetic spin chains in an optical lattice},\ }\href {https://doi.org/10.1038/nature09994} {\bibfield  {journal} {\bibinfo  {journal} {Nature}\ }\textbf {\bibinfo {volume} {472}},\ \bibinfo {pages} {307} (\bibinfo {year} {2011})}\BibitemShut {NoStop}%
\bibitem [{\citenamefont {Tan}\ \emph {et~al.}(2021)\citenamefont {Tan}, \citenamefont {Becker}, \citenamefont {Liu}, \citenamefont {Pagano}, \citenamefont {Collins}, \citenamefont {De}, \citenamefont {Feng}, \citenamefont {Kaplan}, \citenamefont {Kyprianidis}, \citenamefont {Lundgren}, \citenamefont {Morong}, \citenamefont {Whitsitt}, \citenamefont {Gorshkov},\ and\ \citenamefont {Monroe}}]{Tan2021}%
  \BibitemOpen
  \bibfield  {author} {\bibinfo {author} {\bibfnamefont {W.~L.}\ \bibnamefont {Tan}}, \bibinfo {author} {\bibfnamefont {P.}~\bibnamefont {Becker}}, \bibinfo {author} {\bibfnamefont {F.}~\bibnamefont {Liu}}, \bibinfo {author} {\bibfnamefont {G.}~\bibnamefont {Pagano}}, \bibinfo {author} {\bibfnamefont {K.~S.}\ \bibnamefont {Collins}}, \bibinfo {author} {\bibfnamefont {A.}~\bibnamefont {De}}, \bibinfo {author} {\bibfnamefont {L.}~\bibnamefont {Feng}}, \bibinfo {author} {\bibfnamefont {H.~B.}\ \bibnamefont {Kaplan}}, \bibinfo {author} {\bibfnamefont {A.}~\bibnamefont {Kyprianidis}}, \bibinfo {author} {\bibfnamefont {R.}~\bibnamefont {Lundgren}}, \bibinfo {author} {\bibfnamefont {W.}~\bibnamefont {Morong}}, \bibinfo {author} {\bibfnamefont {S.}~\bibnamefont {Whitsitt}}, \bibinfo {author} {\bibfnamefont {A.~V.}\ \bibnamefont {Gorshkov}},\ and\ \bibinfo {author} {\bibfnamefont {C.}~\bibnamefont {Monroe}},\ }\bibfield  {title} {\bibinfo {title} {Domain-wall confinement and dynamics in a quantum simulator},\ }\href
  {https://doi.org/10.1038/s41567-021-01194-3} {\bibfield  {journal} {\bibinfo  {journal} {Nature Physics}\ }\textbf {\bibinfo {volume} {17}},\ \bibinfo {pages} {742} (\bibinfo {year} {2021})}\BibitemShut {NoStop}%
\bibitem [{\citenamefont {Vovrosh}\ and\ \citenamefont {Knolle}(2021)}]{Knolle_2021}%
  \BibitemOpen
  \bibfield  {author} {\bibinfo {author} {\bibfnamefont {J.}~\bibnamefont {Vovrosh}}\ and\ \bibinfo {author} {\bibfnamefont {J.}~\bibnamefont {Knolle}},\ }\bibfield  {title} {\bibinfo {title} {Confinement and entanglement dynamics on a digital quantum computer},\ }\href {https://doi.org/10.1038/s41598-021-90849-5} {\bibfield  {journal} {\bibinfo  {journal} {Scientific Reports}\ }\textbf {\bibinfo {volume} {11}},\ \bibinfo {pages} {11577} (\bibinfo {year} {2021})}\BibitemShut {NoStop}%
\bibitem [{\citenamefont {De}\ \emph {et~al.}(2024)\citenamefont {De}, \citenamefont {Lerose}, \citenamefont {Luo}, \citenamefont {Surace}, \citenamefont {Schuckert}, \citenamefont {Bennewitz}, \citenamefont {Ware}, \citenamefont {Morong}, \citenamefont {Collins}, \citenamefont {Davoudi}, \citenamefont {Gorshkov}, \citenamefont {Katz},\ and\ \citenamefont {Monroe}}]{de2024}%
  \BibitemOpen
  \bibfield  {author} {\bibinfo {author} {\bibfnamefont {A.}~\bibnamefont {De}}, \bibinfo {author} {\bibfnamefont {A.}~\bibnamefont {Lerose}}, \bibinfo {author} {\bibfnamefont {D.}~\bibnamefont {Luo}}, \bibinfo {author} {\bibfnamefont {F.~M.}\ \bibnamefont {Surace}}, \bibinfo {author} {\bibfnamefont {A.}~\bibnamefont {Schuckert}}, \bibinfo {author} {\bibfnamefont {E.~R.}\ \bibnamefont {Bennewitz}}, \bibinfo {author} {\bibfnamefont {B.}~\bibnamefont {Ware}}, \bibinfo {author} {\bibfnamefont {W.}~\bibnamefont {Morong}}, \bibinfo {author} {\bibfnamefont {K.~S.}\ \bibnamefont {Collins}}, \bibinfo {author} {\bibfnamefont {Z.}~\bibnamefont {Davoudi}}, \bibinfo {author} {\bibfnamefont {A.~V.}\ \bibnamefont {Gorshkov}}, \bibinfo {author} {\bibfnamefont {O.}~\bibnamefont {Katz}},\ and\ \bibinfo {author} {\bibfnamefont {C.}~\bibnamefont {Monroe}},\ }\href {https://arxiv.org/abs/2410.13815} {\bibinfo {title} {Observation of string-breaking dynamics in a quantum simulator}} (\bibinfo {year} {2024}),\ \Eprint
  {https://arxiv.org/abs/2410.13815} {arXiv:2410.13815 [quant-ph]} \BibitemShut {NoStop}%
\bibitem [{\citenamefont {Mildenberger}\ \emph {et~al.}(2025)\citenamefont {Mildenberger}, \citenamefont {Mruczkiewicz}, \citenamefont {Halimeh}, \citenamefont {Jiang},\ and\ \citenamefont {Hauke}}]{Mildenberger2025}%
  \BibitemOpen
  \bibfield  {author} {\bibinfo {author} {\bibfnamefont {J.}~\bibnamefont {Mildenberger}}, \bibinfo {author} {\bibfnamefont {W.}~\bibnamefont {Mruczkiewicz}}, \bibinfo {author} {\bibfnamefont {J.~C.}\ \bibnamefont {Halimeh}}, \bibinfo {author} {\bibfnamefont {Z.}~\bibnamefont {Jiang}},\ and\ \bibinfo {author} {\bibfnamefont {P.}~\bibnamefont {Hauke}},\ }\bibfield  {title} {\bibinfo {title} {Confinement in a ${{\mathbb{Z}}}_{2}$ lattice gauge theory on a quantum computer},\ }\href {https://doi.org/10.1038/s41567-024-02723-6} {\bibfield  {journal} {\bibinfo  {journal} {Nature Physics}\ }\textbf {\bibinfo {volume} {21}},\ \bibinfo {pages} {312–317} (\bibinfo {year} {2025})}\BibitemShut {NoStop}%
\bibitem [{\citenamefont {Lüscher}\ \emph {et~al.}(1981)\citenamefont {Lüscher}, \citenamefont {Münster},\ and\ \citenamefont {Weisz}}]{LUSCHER_1981_expansion}%
  \BibitemOpen
  \bibfield  {author} {\bibinfo {author} {\bibfnamefont {M.}~\bibnamefont {Lüscher}}, \bibinfo {author} {\bibfnamefont {G.}~\bibnamefont {Münster}},\ and\ \bibinfo {author} {\bibfnamefont {P.}~\bibnamefont {Weisz}},\ }\bibfield  {title} {\bibinfo {title} {How thick are chromo-electric flux tubes?},\ }\href {https://doi.org/https://doi.org/10.1016/0550-3213(81)90151-6} {\bibfield  {journal} {\bibinfo  {journal} {Nuclear Physics B}\ }\textbf {\bibinfo {volume} {180}},\ \bibinfo {pages} {1} (\bibinfo {year} {1981})}\BibitemShut {NoStop}%
\bibitem [{\citenamefont {Lüscher}(1981)}]{LUSCHER_1981_SSB_aspect}%
  \BibitemOpen
  \bibfield  {author} {\bibinfo {author} {\bibfnamefont {M.}~\bibnamefont {Lüscher}},\ }\bibfield  {title} {\bibinfo {title} {Symmetry-breaking aspects of the roughening transition in gauge theories},\ }\href {https://doi.org/https://doi.org/10.1016/0550-3213(81)90423-5} {\bibfield  {journal} {\bibinfo  {journal} {Nuclear Physics B}\ }\textbf {\bibinfo {volume} {180}},\ \bibinfo {pages} {317} (\bibinfo {year} {1981})}\BibitemShut {NoStop}%
\bibitem [{\citenamefont {Hasenfratz}\ \emph {et~al.}(1981)\citenamefont {Hasenfratz}, \citenamefont {Hasenfratz},\ and\ \citenamefont {Hasenfratz}}]{HASENFRATZ_1981_SOS}%
  \BibitemOpen
  \bibfield  {author} {\bibinfo {author} {\bibfnamefont {A.}~\bibnamefont {Hasenfratz}}, \bibinfo {author} {\bibfnamefont {E.}~\bibnamefont {Hasenfratz}},\ and\ \bibinfo {author} {\bibfnamefont {P.}~\bibnamefont {Hasenfratz}},\ }\bibfield  {title} {\bibinfo {title} {Generalized roughening transition and its effect on the string tension},\ }\href {https://doi.org/https://doi.org/10.1016/0550-3213(81)90426-0} {\bibfield  {journal} {\bibinfo  {journal} {Nuclear Physics B}\ }\textbf {\bibinfo {volume} {180}},\ \bibinfo {pages} {353} (\bibinfo {year} {1981})}\BibitemShut {NoStop}%
\bibitem [{\citenamefont {Cochran}\ \emph {et~al.}(2025)\citenamefont {Cochran}, \citenamefont {Jobst}, \citenamefont {Rosenberg}, \citenamefont {Lensky}, \citenamefont {Gyawali} \emph {et~al.}}]{visualizing_2024}%
  \BibitemOpen
  \bibfield  {author} {\bibinfo {author} {\bibfnamefont {T.~A.}\ \bibnamefont {Cochran}}, \bibinfo {author} {\bibfnamefont {B.}~\bibnamefont {Jobst}}, \bibinfo {author} {\bibfnamefont {E.}~\bibnamefont {Rosenberg}}, \bibinfo {author} {\bibfnamefont {Y.~D.}\ \bibnamefont {Lensky}}, \bibinfo {author} {\bibfnamefont {G.}~\bibnamefont {Gyawali}}, \emph {et~al.},\ }\bibfield  {title} {\bibinfo {title} {Visualizing dynamics of charges and strings in (2 + 1)d lattice gauge theories},\ }\href {https://doi.org/10.1038/s41586-025-08999-9} {\bibfield  {journal} {\bibinfo  {journal} {Nature}\ }\textbf {\bibinfo {volume} {642}},\ \bibinfo {pages} {315} (\bibinfo {year} {2025})}\BibitemShut {NoStop}%
\bibitem [{\citenamefont {Caselle}\ \emph {et~al.}(1996)\citenamefont {Caselle}, \citenamefont {Gliozzi}, \citenamefont {Magnea},\ and\ \citenamefont {Vinti}}]{numerical_width_logrithmic_1996}%
  \BibitemOpen
  \bibfield  {author} {\bibinfo {author} {\bibfnamefont {M.}~\bibnamefont {Caselle}}, \bibinfo {author} {\bibfnamefont {F.}~\bibnamefont {Gliozzi}}, \bibinfo {author} {\bibfnamefont {U.}~\bibnamefont {Magnea}},\ and\ \bibinfo {author} {\bibfnamefont {S.}~\bibnamefont {Vinti}},\ }\bibfield  {title} {\bibinfo {title} {Width of long colour flux tubes in lattice gauge systems},\ }\href {https://doi.org/https://doi.org/10.1016/0550-3213(95)00639-7} {\bibfield  {journal} {\bibinfo  {journal} {Nuclear Physics B}\ }\textbf {\bibinfo {volume} {460}},\ \bibinfo {pages} {397} (\bibinfo {year} {1996})}\BibitemShut {NoStop}%
\bibitem [{\citenamefont {Münster}\ and\ \citenamefont {Weisz}(1981)}]{MUNSTER_1981_expansion}%
  \BibitemOpen
  \bibfield  {author} {\bibinfo {author} {\bibfnamefont {G.}~\bibnamefont {Münster}}\ and\ \bibinfo {author} {\bibfnamefont {P.}~\bibnamefont {Weisz}},\ }\bibfield  {title} {\bibinfo {title} {On the roughening transition in abelian lattice gauge theories},\ }\href {https://doi.org/https://doi.org/10.1016/0550-3213(81)90152-8} {\bibfield  {journal} {\bibinfo  {journal} {Nuclear Physics B}\ }\textbf {\bibinfo {volume} {180}},\ \bibinfo {pages} {13} (\bibinfo {year} {1981})}\BibitemShut {NoStop}%
\bibitem [{\citenamefont {Drouffe}\ and\ \citenamefont {Zuber}(1981{\natexlab{a}})}]{DROUFFE_1981}%
  \BibitemOpen
  \bibfield  {author} {\bibinfo {author} {\bibfnamefont {J.}~\bibnamefont {Drouffe}}\ and\ \bibinfo {author} {\bibfnamefont {J.}~\bibnamefont {Zuber}},\ }\bibfield  {title} {\bibinfo {title} {Roughening transition in lattice gauge theories in arbitrary dimension: (i). the z2 case},\ }\href {https://doi.org/https://doi.org/10.1016/0550-3213(81)90418-1} {\bibfield  {journal} {\bibinfo  {journal} {Nuclear Physics B}\ }\textbf {\bibinfo {volume} {180}},\ \bibinfo {pages} {253} (\bibinfo {year} {1981}{\natexlab{a}})}\BibitemShut {NoStop}%
\bibitem [{\citenamefont {Drouffe}\ and\ \citenamefont {Zuber}(1981{\natexlab{b}})}]{DROUFFE_1981_2}%
  \BibitemOpen
  \bibfield  {author} {\bibinfo {author} {\bibfnamefont {J.}~\bibnamefont {Drouffe}}\ and\ \bibinfo {author} {\bibfnamefont {J.}~\bibnamefont {Zuber}},\ }\bibfield  {title} {\bibinfo {title} {Roughening transition in lattice gauge theories in arbitrary dimension: (ii). the groups z3, u(1), su(2), su(3)},\ }\href {https://doi.org/https://doi.org/10.1016/0550-3213(81)90419-3} {\bibfield  {journal} {\bibinfo  {journal} {Nuclear Physics B}\ }\textbf {\bibinfo {volume} {180}},\ \bibinfo {pages} {264} (\bibinfo {year} {1981}{\natexlab{b}})}\BibitemShut {NoStop}%
\bibitem [{\citenamefont {Fradkin}\ and\ \citenamefont {Shenker}(1979)}]{Fradkin-Shenker}%
  \BibitemOpen
  \bibfield  {author} {\bibinfo {author} {\bibfnamefont {E.}~\bibnamefont {Fradkin}}\ and\ \bibinfo {author} {\bibfnamefont {S.~H.}\ \bibnamefont {Shenker}},\ }\bibfield  {title} {\bibinfo {title} {Phase diagrams of lattice gauge theories with higgs fields},\ }\href {https://doi.org/10.1103/PhysRevD.19.3682} {\bibfield  {journal} {\bibinfo  {journal} {Phys. Rev. D}\ }\textbf {\bibinfo {volume} {19}},\ \bibinfo {pages} {3682} (\bibinfo {year} {1979})}\BibitemShut {NoStop}%
\bibitem [{\citenamefont {Wegner}(1971)}]{Wegner_duality_1971}%
  \BibitemOpen
  \bibfield  {author} {\bibinfo {author} {\bibfnamefont {F.~J.}\ \bibnamefont {Wegner}},\ }\bibfield  {title} {\bibinfo {title} {Duality in generalized ising models and phase transitions without local order parameters},\ }\href {https://pubs.aip.org/aip/jmp/article-abstract/12/10/2259/465334/Duality-in-Generalized-Ising-Models-and-Phase?redirectedFrom=fulltext} {\bibfield  {journal} {\bibinfo  {journal} {Journal of Mathematical Physics}\ }\textbf {\bibinfo {volume} {12}},\ \bibinfo {pages} {2259} (\bibinfo {year} {1971})}\BibitemShut {NoStop}%
\bibitem [{\citenamefont {Fradkin}(1983)}]{Fradkin_1983}%
  \BibitemOpen
  \bibfield  {author} {\bibinfo {author} {\bibfnamefont {E.}~\bibnamefont {Fradkin}},\ }\bibfield  {title} {\bibinfo {title} {Roughening transition in quantum interfaces},\ }\href {https://doi.org/10.1103/PhysRevB.28.5338} {\bibfield  {journal} {\bibinfo  {journal} {Phys. Rev. B}\ }\textbf {\bibinfo {volume} {28}},\ \bibinfo {pages} {5338} (\bibinfo {year} {1983})}\BibitemShut {NoStop}%
\bibitem [{\citenamefont {B{\"u}rkner}\ and\ \citenamefont {Stauffer}(1983)}]{Burkner_1983_MC_Ising}%
  \BibitemOpen
  \bibfield  {author} {\bibinfo {author} {\bibfnamefont {E.}~\bibnamefont {B{\"u}rkner}}\ and\ \bibinfo {author} {\bibfnamefont {D.}~\bibnamefont {Stauffer}},\ }\bibfield  {title} {\bibinfo {title} {Monte carlo study of surface roughening in the three-dimensional ising model},\ }\href {https://doi.org/10.1007/BF01388545} {\bibfield  {journal} {\bibinfo  {journal} {Zeitschrift f{\"u}r Physik B Condensed Matter}\ }\textbf {\bibinfo {volume} {53}},\ \bibinfo {pages} {241} (\bibinfo {year} {1983})}\BibitemShut {NoStop}%
\bibitem [{\citenamefont {Hasenbusch}\ and\ \citenamefont {Pinn}(1997)}]{Matching_Ising_sos_1997}%
  \BibitemOpen
  \bibfield  {author} {\bibinfo {author} {\bibfnamefont {M.}~\bibnamefont {Hasenbusch}}\ and\ \bibinfo {author} {\bibfnamefont {K.}~\bibnamefont {Pinn}},\ }\bibfield  {title} {\bibinfo {title} {Computing the roughening transition of ising and solid-on-solid models by bcsos model matching},\ }\href {https://doi.org/10.1088/0305-4470/30/1/006} {\bibfield  {journal} {\bibinfo  {journal} {Journal of Physics A: Mathematical and General}\ }\textbf {\bibinfo {volume} {30}},\ \bibinfo {pages} {63} (\bibinfo {year} {1997})}\BibitemShut {NoStop}%
\bibitem [{\citenamefont {Caselle}\ \emph {et~al.}(2003)\citenamefont {Caselle}, \citenamefont {Hasenbusch},\ and\ \citenamefont {Panero}}]{MC_Ising_interface_2003}%
  \BibitemOpen
  \bibfield  {author} {\bibinfo {author} {\bibfnamefont {M.}~\bibnamefont {Caselle}}, \bibinfo {author} {\bibfnamefont {M.}~\bibnamefont {Hasenbusch}},\ and\ \bibinfo {author} {\bibfnamefont {M.}~\bibnamefont {Panero}},\ }\bibfield  {title} {\bibinfo {title} {String effects in the 3d gauge ising model},\ }\href {https://doi.org/10.1088/1126-6708/2003/01/057} {\bibfield  {journal} {\bibinfo  {journal} {Journal of High Energy Physics}\ }\textbf {\bibinfo {volume} {2003}},\ \bibinfo {pages} {057} (\bibinfo {year} {2003})}\BibitemShut {NoStop}%
\bibitem [{\citenamefont {Caselle}\ \emph {et~al.}(2007)\citenamefont {Caselle}, \citenamefont {Hasenbusch},\ and\ \citenamefont {Panero}}]{3d_Ising_interface_Nabum_Goto_2007}%
  \BibitemOpen
  \bibfield  {author} {\bibinfo {author} {\bibfnamefont {M.}~\bibnamefont {Caselle}}, \bibinfo {author} {\bibfnamefont {M.}~\bibnamefont {Hasenbusch}},\ and\ \bibinfo {author} {\bibfnamefont {M.}~\bibnamefont {Panero}},\ }\bibfield  {title} {\bibinfo {title} {The interface free energy: comparison of accurate monte carlo results for the 3d ising model with effective interface models},\ }\href {https://doi.org/10.1088/1126-6708/2007/09/117} {\bibfield  {journal} {\bibinfo  {journal} {Journal of High Energy Physics}\ }\textbf {\bibinfo {volume} {2007}},\ \bibinfo {pages} {117} (\bibinfo {year} {2007})}\BibitemShut {NoStop}%
\bibitem [{\citenamefont {Dashti-Naserabadi}\ \emph {et~al.}(2019)\citenamefont {Dashti-Naserabadi}, \citenamefont {Saberi}, \citenamefont {Rahbari},\ and\ \citenamefont {Park}}]{Roughen_3D_classical_Ising_2019}%
  \BibitemOpen
  \bibfield  {author} {\bibinfo {author} {\bibfnamefont {H.}~\bibnamefont {Dashti-Naserabadi}}, \bibinfo {author} {\bibfnamefont {A.~A.}\ \bibnamefont {Saberi}}, \bibinfo {author} {\bibfnamefont {S.~H.~E.}\ \bibnamefont {Rahbari}},\ and\ \bibinfo {author} {\bibfnamefont {H.}~\bibnamefont {Park}},\ }\bibfield  {title} {\bibinfo {title} {Two-dimensional super-roughening in the three-dimensional ising model},\ }\href {https://doi.org/10.1103/PhysRevE.100.060101} {\bibfield  {journal} {\bibinfo  {journal} {Phys. Rev. E}\ }\textbf {\bibinfo {volume} {100}},\ \bibinfo {pages} {060101} (\bibinfo {year} {2019})}\BibitemShut {NoStop}%
\bibitem [{\citenamefont {Krinitsin}\ \emph {et~al.}(2024)\citenamefont {Krinitsin}, \citenamefont {Tausendpfund}, \citenamefont {Rizzi}, \citenamefont {Heyl},\ and\ \citenamefont {Schmitt}}]{roughening_dynamics_2024}%
  \BibitemOpen
  \bibfield  {author} {\bibinfo {author} {\bibfnamefont {W.}~\bibnamefont {Krinitsin}}, \bibinfo {author} {\bibfnamefont {N.}~\bibnamefont {Tausendpfund}}, \bibinfo {author} {\bibfnamefont {M.}~\bibnamefont {Rizzi}}, \bibinfo {author} {\bibfnamefont {M.}~\bibnamefont {Heyl}},\ and\ \bibinfo {author} {\bibfnamefont {M.}~\bibnamefont {Schmitt}},\ }\href {https://arxiv.org/abs/2412.10145} {\bibinfo {title} {Roughening dynamics of interfaces in two-dimensional quantum matter}} (\bibinfo {year} {2024}),\ \Eprint {https://arxiv.org/abs/2412.10145} {arXiv:2412.10145 [quant-ph]} \BibitemShut {NoStop}%
\bibitem [{\citenamefont {Kosterlitz}\ and\ \citenamefont {Thouless}(1973)}]{KT_1973}%
  \BibitemOpen
  \bibfield  {author} {\bibinfo {author} {\bibfnamefont {J.~M.}\ \bibnamefont {Kosterlitz}}\ and\ \bibinfo {author} {\bibfnamefont {D.~J.}\ \bibnamefont {Thouless}},\ }\bibfield  {title} {\bibinfo {title} {Ordering, metastability and phase transitions in two-dimensional systems},\ }\href {https://doi.org/10.1088/0022-3719/6/7/010} {\bibfield  {journal} {\bibinfo  {journal} {Journal of Physics C: Solid State Physics}\ }\textbf {\bibinfo {volume} {6}},\ \bibinfo {pages} {1181} (\bibinfo {year} {1973})}\BibitemShut {NoStop}%
\bibitem [{\citenamefont {{Berezinski{\v{i}}}}(1971)}]{Berezinskii_1971}%
  \BibitemOpen
  \bibfield  {author} {\bibinfo {author} {\bibfnamefont {V.~L.}\ \bibnamefont {{Berezinski{\v{i}}}}},\ }\bibfield  {title} {\bibinfo {title} {{Destruction of Long-range Order in One-dimensional and Two-dimensional Systems having a Continuous Symmetry Group I. Classical Systems}},\ }\href@noop {} {\bibfield  {journal} {\bibinfo  {journal} {Soviet Journal of Experimental and Theoretical Physics}\ }\textbf {\bibinfo {volume} {32}},\ \bibinfo {pages} {493} (\bibinfo {year} {1971})}\BibitemShut {NoStop}%
\bibitem [{\citenamefont {{Berezinski{\v{i}}}}(1972)}]{Berezinskii_1972}%
  \BibitemOpen
  \bibfield  {author} {\bibinfo {author} {\bibfnamefont {V.~L.}\ \bibnamefont {{Berezinski{\v{i}}}}},\ }\bibfield  {title} {\bibinfo {title} {{Destruction of Long-range Order in One-dimensional and Two-dimensional Systems Possessing a Continuous Symmetry Group. II. Quantum Systems}},\ }\href@noop {} {\bibfield  {journal} {\bibinfo  {journal} {Soviet Journal of Experimental and Theoretical Physics}\ }\textbf {\bibinfo {volume} {34}},\ \bibinfo {pages} {610} (\bibinfo {year} {1972})}\BibitemShut {NoStop}%
\bibitem [{\citenamefont {Batista}\ and\ \citenamefont {Nussinov}(2005)}]{Zohar_2004}%
  \BibitemOpen
  \bibfield  {author} {\bibinfo {author} {\bibfnamefont {C.~D.}\ \bibnamefont {Batista}}\ and\ \bibinfo {author} {\bibfnamefont {Z.}~\bibnamefont {Nussinov}},\ }\bibfield  {title} {\bibinfo {title} {Generalized elitzur's theorem and dimensional reductions},\ }\href {https://doi.org/10.1103/PhysRevB.72.045137} {\bibfield  {journal} {\bibinfo  {journal} {Phys. Rev. B}\ }\textbf {\bibinfo {volume} {72}},\ \bibinfo {pages} {045137} (\bibinfo {year} {2005})}\BibitemShut {NoStop}%
\bibitem [{\citenamefont {Nussinov}\ and\ \citenamefont {Ortiz}(2009)}]{NUSSINOV_2009}%
  \BibitemOpen
  \bibfield  {author} {\bibinfo {author} {\bibfnamefont {Z.}~\bibnamefont {Nussinov}}\ and\ \bibinfo {author} {\bibfnamefont {G.}~\bibnamefont {Ortiz}},\ }\bibfield  {title} {\bibinfo {title} {A symmetry principle for topological quantum order},\ }\href {https://doi.org/https://doi.org/10.1016/j.aop.2008.11.002} {\bibfield  {journal} {\bibinfo  {journal} {Annals of Physics}\ }\textbf {\bibinfo {volume} {324}},\ \bibinfo {pages} {977} (\bibinfo {year} {2009})}\BibitemShut {NoStop}%
\bibitem [{\citenamefont {Gaiotto}\ \emph {et~al.}(2015)\citenamefont {Gaiotto}, \citenamefont {Kapustin}, \citenamefont {Seiberg},\ and\ \citenamefont {Willett}}]{High_form_Kapistin_2015}%
  \BibitemOpen
  \bibfield  {author} {\bibinfo {author} {\bibfnamefont {D.}~\bibnamefont {Gaiotto}}, \bibinfo {author} {\bibfnamefont {A.}~\bibnamefont {Kapustin}}, \bibinfo {author} {\bibfnamefont {N.}~\bibnamefont {Seiberg}},\ and\ \bibinfo {author} {\bibfnamefont {B.}~\bibnamefont {Willett}},\ }\bibfield  {title} {\bibinfo {title} {Generalized global symmetries},\ }\href {https://link-springer-com.eaccess.tum.edu/article/10.1007/JHEP02(2015)172} {\bibfield  {journal} {\bibinfo  {journal} {Journal of High Energy Physics}\ }\textbf {\bibinfo {volume} {2015}},\ \bibinfo {pages} {1} (\bibinfo {year} {2015})}\BibitemShut {NoStop}%
\bibitem [{\citenamefont {McGreevy}(2023)}]{McGreevy_2023}%
  \BibitemOpen
  \bibfield  {author} {\bibinfo {author} {\bibfnamefont {J.}~\bibnamefont {McGreevy}},\ }\bibfield  {title} {\bibinfo {title} {Generalized symmetries in condensed matter},\ }\href {https://doi.org/10.1146/annurev-conmatphys-040721-021029} {\bibfield  {journal} {\bibinfo  {journal} {Annual Review of Condensed Matter Physics}\ }\textbf {\bibinfo {volume} {14}},\ \bibinfo {pages} {57} (\bibinfo {year} {2023})},\ \Eprint {https://arxiv.org/abs/https://doi.org/10.1146/annurev-conmatphys-040721-021029} {https://doi.org/10.1146/annurev-conmatphys-040721-021029} \BibitemShut {NoStop}%
\bibitem [{\citenamefont {Bhardwaj}\ \emph {et~al.}(2024)\citenamefont {Bhardwaj}, \citenamefont {Bottini}, \citenamefont {Fraser-Taliente}, \citenamefont {Gladden}, \citenamefont {Gould}, \citenamefont {Platschorre},\ and\ \citenamefont {Tillim}}]{Bhardwaj_2023}%
  \BibitemOpen
  \bibfield  {author} {\bibinfo {author} {\bibfnamefont {L.}~\bibnamefont {Bhardwaj}}, \bibinfo {author} {\bibfnamefont {L.~E.}\ \bibnamefont {Bottini}}, \bibinfo {author} {\bibfnamefont {L.}~\bibnamefont {Fraser-Taliente}}, \bibinfo {author} {\bibfnamefont {L.}~\bibnamefont {Gladden}}, \bibinfo {author} {\bibfnamefont {D.~S.~W.}\ \bibnamefont {Gould}}, \bibinfo {author} {\bibfnamefont {A.}~\bibnamefont {Platschorre}},\ and\ \bibinfo {author} {\bibfnamefont {H.}~\bibnamefont {Tillim}},\ }\bibfield  {title} {\bibinfo {title} {{Lectures on generalized symmetries}},\ }\href {https://doi.org/10.1016/j.physrep.2023.11.002} {\bibfield  {journal} {\bibinfo  {journal} {Phys. Rept.}\ }\textbf {\bibinfo {volume} {1051}},\ \bibinfo {pages} {1} (\bibinfo {year} {2024})},\ \Eprint {https://arxiv.org/abs/2307.07547} {arXiv:2307.07547 [hep-th]} \BibitemShut {NoStop}%
\bibitem [{\citenamefont {Hastings}\ and\ \citenamefont {Wen}(2005)}]{Hastings_and_Wen_2005}%
  \BibitemOpen
  \bibfield  {author} {\bibinfo {author} {\bibfnamefont {M.~B.}\ \bibnamefont {Hastings}}\ and\ \bibinfo {author} {\bibfnamefont {X.-G.}\ \bibnamefont {Wen}},\ }\bibfield  {title} {\bibinfo {title} {Quasiadiabatic continuation of quantum states: The stability of topological ground-state degeneracy and emergent gauge invariance},\ }\href {https://doi.org/10.1103/PhysRevB.72.045141} {\bibfield  {journal} {\bibinfo  {journal} {Phys. Rev. B}\ }\textbf {\bibinfo {volume} {72}},\ \bibinfo {pages} {045141} (\bibinfo {year} {2005})}\BibitemShut {NoStop}%
\bibitem [{\citenamefont {Wen}(2019)}]{High_form_wen_2019}%
  \BibitemOpen
  \bibfield  {author} {\bibinfo {author} {\bibfnamefont {X.-G.}\ \bibnamefont {Wen}},\ }\bibfield  {title} {\bibinfo {title} {Emergent anomalous higher symmetries from topological order and from dynamical electromagnetic field in condensed matter systems},\ }\href {https://doi.org/10.1103/PhysRevB.99.205139} {\bibfield  {journal} {\bibinfo  {journal} {Phys. Rev. B}\ }\textbf {\bibinfo {volume} {99}},\ \bibinfo {pages} {205139} (\bibinfo {year} {2019})}\BibitemShut {NoStop}%
\bibitem [{\citenamefont {Somoza}\ \emph {et~al.}(2021)\citenamefont {Somoza}, \citenamefont {Serna},\ and\ \citenamefont {Nahum}}]{Adam_Nahum_2021}%
  \BibitemOpen
  \bibfield  {author} {\bibinfo {author} {\bibfnamefont {A.~M.}\ \bibnamefont {Somoza}}, \bibinfo {author} {\bibfnamefont {P.}~\bibnamefont {Serna}},\ and\ \bibinfo {author} {\bibfnamefont {A.}~\bibnamefont {Nahum}},\ }\bibfield  {title} {\bibinfo {title} {Self-dual criticality in three-dimensional $\mathbb{Z}_{2}$ gauge theory with matter},\ }\href {https://doi.org/10.1103/PhysRevX.11.041008} {\bibfield  {journal} {\bibinfo  {journal} {Phys. Rev. X}\ }\textbf {\bibinfo {volume} {11}},\ \bibinfo {pages} {041008} (\bibinfo {year} {2021})}\BibitemShut {NoStop}%
\bibitem [{\citenamefont {Pace}\ and\ \citenamefont {Wen}(2023)}]{Wen_emergent_high_form_2023}%
  \BibitemOpen
  \bibfield  {author} {\bibinfo {author} {\bibfnamefont {S.~D.}\ \bibnamefont {Pace}}\ and\ \bibinfo {author} {\bibfnamefont {X.-G.}\ \bibnamefont {Wen}},\ }\bibfield  {title} {\bibinfo {title} {Exact emergent higher-form symmetries in bosonic lattice models},\ }\href {https://doi.org/10.1103/PhysRevB.108.195147} {\bibfield  {journal} {\bibinfo  {journal} {Phys. Rev. B}\ }\textbf {\bibinfo {volume} {108}},\ \bibinfo {pages} {195147} (\bibinfo {year} {2023})}\BibitemShut {NoStop}%
\bibitem [{\citenamefont {Cherman}\ and\ \citenamefont {Jacobson}(2024)}]{Emergent_1_form_PRD_2024}%
  \BibitemOpen
  \bibfield  {author} {\bibinfo {author} {\bibfnamefont {A.}~\bibnamefont {Cherman}}\ and\ \bibinfo {author} {\bibfnamefont {T.}~\bibnamefont {Jacobson}},\ }\bibfield  {title} {\bibinfo {title} {Emergent 1-form symmetries},\ }\href {https://doi.org/10.1103/PhysRevD.109.125013} {\bibfield  {journal} {\bibinfo  {journal} {Phys. Rev. D}\ }\textbf {\bibinfo {volume} {109}},\ \bibinfo {pages} {125013} (\bibinfo {year} {2024})}\BibitemShut {NoStop}%
\bibitem [{\citenamefont {Serna}\ \emph {et~al.}(2024)\citenamefont {Serna}, \citenamefont {Somoza},\ and\ \citenamefont {Nahum}}]{Adam_Nahum_2024}%
  \BibitemOpen
  \bibfield  {author} {\bibinfo {author} {\bibfnamefont {P.}~\bibnamefont {Serna}}, \bibinfo {author} {\bibfnamefont {A.~M.}\ \bibnamefont {Somoza}},\ and\ \bibinfo {author} {\bibfnamefont {A.}~\bibnamefont {Nahum}},\ }\bibfield  {title} {\bibinfo {title} {Worldsheet patching, 1-form symmetries, and ${\mathrm{landau}}^{*}$ phase transitions},\ }\href {https://doi.org/10.1103/PhysRevB.110.115102} {\bibfield  {journal} {\bibinfo  {journal} {Phys. Rev. B}\ }\textbf {\bibinfo {volume} {110}},\ \bibinfo {pages} {115102} (\bibinfo {year} {2024})}\BibitemShut {NoStop}%
\bibitem [{\citenamefont {Liu}\ \emph {et~al.}(2025)\citenamefont {Liu}, \citenamefont {Xu}, \citenamefont {Pollmann},\ and\ \citenamefont {Knap}}]{QEC_1_form_2025}%
  \BibitemOpen
  \bibfield  {author} {\bibinfo {author} {\bibfnamefont {Y.-J.}\ \bibnamefont {Liu}}, \bibinfo {author} {\bibfnamefont {W.-T.}\ \bibnamefont {Xu}}, \bibinfo {author} {\bibfnamefont {F.}~\bibnamefont {Pollmann}},\ and\ \bibinfo {author} {\bibfnamefont {M.}~\bibnamefont {Knap}},\ }\href {https://arxiv.org/abs/2502.17572} {\bibinfo {title} {Detecting emergent 1-form symmetries with quantum error correction}} (\bibinfo {year} {2025}),\ \Eprint {https://arxiv.org/abs/2502.17572} {arXiv:2502.17572 [quant-ph]} \BibitemShut {NoStop}%
\bibitem [{\citenamefont {Kogut}\ and\ \citenamefont {Sinclair}(1981)}]{Kogut_1981_PRD_1}%
  \BibitemOpen
  \bibfield  {author} {\bibinfo {author} {\bibfnamefont {J.~B.}\ \bibnamefont {Kogut}}\ and\ \bibinfo {author} {\bibfnamefont {D.~K.}\ \bibnamefont {Sinclair}},\ }\bibfield  {title} {\bibinfo {title} {Analyticity of the off-axis string tension and the restoration of rotational symmetry in lattice systems},\ }\href {https://doi.org/10.1103/PhysRevD.24.1610} {\bibfield  {journal} {\bibinfo  {journal} {Phys. Rev. D}\ }\textbf {\bibinfo {volume} {24}},\ \bibinfo {pages} {1610} (\bibinfo {year} {1981})}\BibitemShut {NoStop}%
\bibitem [{\citenamefont {Kogut}\ \emph {et~al.}(1981)\citenamefont {Kogut}, \citenamefont {Sinclair}, \citenamefont {Pearson}, \citenamefont {Richardson},\ and\ \citenamefont {Shigemitsu}}]{Kogut_1981_PRD_2}%
  \BibitemOpen
  \bibfield  {author} {\bibinfo {author} {\bibfnamefont {J.~B.}\ \bibnamefont {Kogut}}, \bibinfo {author} {\bibfnamefont {D.~K.}\ \bibnamefont {Sinclair}}, \bibinfo {author} {\bibfnamefont {R.~B.}\ \bibnamefont {Pearson}}, \bibinfo {author} {\bibfnamefont {J.~L.}\ \bibnamefont {Richardson}},\ and\ \bibinfo {author} {\bibfnamefont {J.}~\bibnamefont {Shigemitsu}},\ }\bibfield  {title} {\bibinfo {title} {Fluctuating string of lattice gauge theory: The heavy-quark potential, the restoration of rotational symmetry, and roughening},\ }\href {https://doi.org/10.1103/PhysRevD.23.2945} {\bibfield  {journal} {\bibinfo  {journal} {Phys. Rev. D}\ }\textbf {\bibinfo {volume} {23}},\ \bibinfo {pages} {2945} (\bibinfo {year} {1981})}\BibitemShut {NoStop}%
\bibitem [{\citenamefont {White}(1992)}]{DMRG_1992}%
  \BibitemOpen
  \bibfield  {author} {\bibinfo {author} {\bibfnamefont {S.~R.}\ \bibnamefont {White}},\ }\bibfield  {title} {\bibinfo {title} {Density matrix formulation for quantum renormalization groups},\ }\href {https://doi.org/10.1103/PhysRevLett.69.2863} {\bibfield  {journal} {\bibinfo  {journal} {Phys. Rev. Lett.}\ }\textbf {\bibinfo {volume} {69}},\ \bibinfo {pages} {2863} (\bibinfo {year} {1992})}\BibitemShut {NoStop}%
\bibitem [{\citenamefont {White}(1993)}]{DMRG_1993}%
  \BibitemOpen
  \bibfield  {author} {\bibinfo {author} {\bibfnamefont {S.~R.}\ \bibnamefont {White}},\ }\bibfield  {title} {\bibinfo {title} {Density-matrix algorithms for quantum renormalization groups},\ }\href {https://doi.org/10.1103/PhysRevB.48.10345} {\bibfield  {journal} {\bibinfo  {journal} {Phys. Rev. B}\ }\textbf {\bibinfo {volume} {48}},\ \bibinfo {pages} {10345} (\bibinfo {year} {1993})}\BibitemShut {NoStop}%
\bibitem [{\citenamefont {McCulloch}(2008)}]{mcculloch_2008_iDMRG}%
  \BibitemOpen
  \bibfield  {author} {\bibinfo {author} {\bibfnamefont {I.~P.}\ \bibnamefont {McCulloch}},\ }\href {https://arxiv.org/abs/0804.2509} {\bibinfo {title} {Infinite size density matrix renormalization group, revisited}} (\bibinfo {year} {2008}),\ \Eprint {https://arxiv.org/abs/0804.2509} {arXiv:0804.2509 [cond-mat.str-el]} \BibitemShut {NoStop}%
\bibitem [{\citenamefont {Schollwöck}(2011)}]{SCHOLLWOCK_2011}%
  \BibitemOpen
  \bibfield  {author} {\bibinfo {author} {\bibfnamefont {U.}~\bibnamefont {Schollwöck}},\ }\bibfield  {title} {\bibinfo {title} {The density-matrix renormalization group in the age of matrix product states},\ }\href {https://doi.org/https://doi.org/10.1016/j.aop.2010.09.012} {\bibfield  {journal} {\bibinfo  {journal} {Annals of Physics}\ }\textbf {\bibinfo {volume} {326}},\ \bibinfo {pages} {96} (\bibinfo {year} {2011})},\ \bibinfo {note} {january 2011 Special Issue}\BibitemShut {NoStop}%
\bibitem [{\citenamefont {Kogut}\ and\ \citenamefont {Susskind}(1975)}]{Kogut_Susskind}%
  \BibitemOpen
  \bibfield  {author} {\bibinfo {author} {\bibfnamefont {J.}~\bibnamefont {Kogut}}\ and\ \bibinfo {author} {\bibfnamefont {L.}~\bibnamefont {Susskind}},\ }\bibfield  {title} {\bibinfo {title} {Hamiltonian formulation of wilson's lattice gauge theories},\ }\href {https://doi.org/10.1103/PhysRevD.11.395} {\bibfield  {journal} {\bibinfo  {journal} {Phys. Rev. D}\ }\textbf {\bibinfo {volume} {11}},\ \bibinfo {pages} {395} (\bibinfo {year} {1975})}\BibitemShut {NoStop}%
\bibitem [{\citenamefont {Kogut}(1979)}]{Kogut_LGH}%
  \BibitemOpen
  \bibfield  {author} {\bibinfo {author} {\bibfnamefont {J.~B.}\ \bibnamefont {Kogut}},\ }\bibfield  {title} {\bibinfo {title} {An introduction to lattice gauge theory and spin systems},\ }\href {https://doi.org/10.1103/RevModPhys.51.659} {\bibfield  {journal} {\bibinfo  {journal} {Rev. Mod. Phys.}\ }\textbf {\bibinfo {volume} {51}},\ \bibinfo {pages} {659} (\bibinfo {year} {1979})}\BibitemShut {NoStop}%
\bibitem [{\citenamefont {Tupitsyn}\ \emph {et~al.}(2010)\citenamefont {Tupitsyn}, \citenamefont {Kitaev}, \citenamefont {Prokof'ev},\ and\ \citenamefont {Stamp}}]{TC_multi_critical_2010}%
  \BibitemOpen
  \bibfield  {author} {\bibinfo {author} {\bibfnamefont {I.~S.}\ \bibnamefont {Tupitsyn}}, \bibinfo {author} {\bibfnamefont {A.}~\bibnamefont {Kitaev}}, \bibinfo {author} {\bibfnamefont {N.~V.}\ \bibnamefont {Prokof'ev}},\ and\ \bibinfo {author} {\bibfnamefont {P.~C.~E.}\ \bibnamefont {Stamp}},\ }\bibfield  {title} {\bibinfo {title} {Topological multicritical point in the phase diagram of the toric code model and three-dimensional lattice gauge higgs model},\ }\href {https://doi.org/10.1103/PhysRevB.82.085114} {\bibfield  {journal} {\bibinfo  {journal} {Phys. Rev. B}\ }\textbf {\bibinfo {volume} {82}},\ \bibinfo {pages} {085114} (\bibinfo {year} {2010})}\BibitemShut {NoStop}%
\bibitem [{\citenamefont {Xu}\ \emph {et~al.}(2025{\natexlab{a}})\citenamefont {Xu}, \citenamefont {Rakovszky}, \citenamefont {Knap},\ and\ \citenamefont {Pollmann}}]{xu_2024_entanglement}%
  \BibitemOpen
  \bibfield  {author} {\bibinfo {author} {\bibfnamefont {W.-T.}\ \bibnamefont {Xu}}, \bibinfo {author} {\bibfnamefont {T.}~\bibnamefont {Rakovszky}}, \bibinfo {author} {\bibfnamefont {M.}~\bibnamefont {Knap}},\ and\ \bibinfo {author} {\bibfnamefont {F.}~\bibnamefont {Pollmann}},\ }\bibfield  {title} {\bibinfo {title} {Entanglement properties of gauge theories from higher-form symmetries},\ }\href {https://doi.org/10.1103/PhysRevX.15.011001} {\bibfield  {journal} {\bibinfo  {journal} {Phys. Rev. X}\ }\textbf {\bibinfo {volume} {15}},\ \bibinfo {pages} {011001} (\bibinfo {year} {2025}{\natexlab{a}})}\BibitemShut {NoStop}%
\bibitem [{\citenamefont {Kitaev}(2003)}]{kitaev_2002}%
  \BibitemOpen
  \bibfield  {author} {\bibinfo {author} {\bibfnamefont {A.}~\bibnamefont {Kitaev}},\ }\bibfield  {title} {\bibinfo {title} {Fault-tolerant quantum computation by anyons},\ }\href {https://doi.org/https://doi.org/10.1016/S0003-4916(02)00018-0} {\bibfield  {journal} {\bibinfo  {journal} {Annals of Physics}\ }\textbf {\bibinfo {volume} {303}},\ \bibinfo {pages} {2} (\bibinfo {year} {2003})}\BibitemShut {NoStop}%
\bibitem [{app()}]{appendix}%
  \BibitemOpen
  \href@noop {} {\bibinfo  {journal} {See Supplemental Material for the roughening transition in the view point of the dual Ising model; the disorder parameter of the translational symmetry and the Wilson loop correlator; details on sequential iDMRG simulations; supplementary data of the $\mathbb{Z}_2$ lattice gauge theory on a cylinder with a circumference $L_y$ = 8; and the results from the $\mathbb{Z}_3$ clock model. The Supplemental Material also contains Refs.~\cite{Trebst_2007,Duality_TN,Fradkin_2017_disorder_para,yellow_book_CFT,Deng_2002}}\ }\BibitemShut {NoStop}%
\bibitem [{\citenamefont {Tagliacozzo}\ \emph {et~al.}(2008)\citenamefont {Tagliacozzo}, \citenamefont {de~Oliveira}, \citenamefont {Iblisdir},\ and\ \citenamefont {Latorre}}]{Luca_2008}%
  \BibitemOpen
\bibfield  {journal} {  }\bibfield  {author} {\bibinfo {author} {\bibfnamefont {L.}~\bibnamefont {Tagliacozzo}}, \bibinfo {author} {\bibfnamefont {T.~R.}\ \bibnamefont {de~Oliveira}}, \bibinfo {author} {\bibfnamefont {S.}~\bibnamefont {Iblisdir}},\ and\ \bibinfo {author} {\bibfnamefont {J.~I.}\ \bibnamefont {Latorre}},\ }\bibfield  {title} {\bibinfo {title} {Scaling of entanglement support for matrix product states},\ }\href {https://doi.org/10.1103/PhysRevB.78.024410} {\bibfield  {journal} {\bibinfo  {journal} {Phys. Rev. B}\ }\textbf {\bibinfo {volume} {78}},\ \bibinfo {pages} {024410} (\bibinfo {year} {2008})}\BibitemShut {NoStop}%
\bibitem [{\citenamefont {Pollmann}\ \emph {et~al.}(2009)\citenamefont {Pollmann}, \citenamefont {Mukerjee}, \citenamefont {Turner},\ and\ \citenamefont {Moore}}]{Pollmann_2009}%
  \BibitemOpen
  \bibfield  {author} {\bibinfo {author} {\bibfnamefont {F.}~\bibnamefont {Pollmann}}, \bibinfo {author} {\bibfnamefont {S.}~\bibnamefont {Mukerjee}}, \bibinfo {author} {\bibfnamefont {A.~M.}\ \bibnamefont {Turner}},\ and\ \bibinfo {author} {\bibfnamefont {J.~E.}\ \bibnamefont {Moore}},\ }\bibfield  {title} {\bibinfo {title} {Theory of finite-entanglement scaling at one-dimensional quantum critical points},\ }\href {https://doi.org/10.1103/PhysRevLett.102.255701} {\bibfield  {journal} {\bibinfo  {journal} {Phys. Rev. Lett.}\ }\textbf {\bibinfo {volume} {102}},\ \bibinfo {pages} {255701} (\bibinfo {year} {2009})}\BibitemShut {NoStop}%
\bibitem [{\citenamefont {Tschirsich}\ \emph {et~al.}(2019)\citenamefont {Tschirsich}, \citenamefont {Montangero},\ and\ \citenamefont {Dalmonte}}]{Marcello_Dalmonte_2019}%
  \BibitemOpen
  \bibfield  {author} {\bibinfo {author} {\bibfnamefont {F.}~\bibnamefont {Tschirsich}}, \bibinfo {author} {\bibfnamefont {S.}~\bibnamefont {Montangero}},\ and\ \bibinfo {author} {\bibfnamefont {M.}~\bibnamefont {Dalmonte}},\ }\bibfield  {title} {\bibinfo {title} {{Phase diagram and conformal string excitations of square ice using gauge invariant matrix product states}},\ }\href {https://doi.org/10.21468/SciPostPhys.6.3.028} {\bibfield  {journal} {\bibinfo  {journal} {SciPost Phys.}\ }\textbf {\bibinfo {volume} {6}},\ \bibinfo {pages} {028} (\bibinfo {year} {2019})}\BibitemShut {NoStop}%
\bibitem [{Note1()}]{Note1}%
  \BibitemOpen
  \bibinfo {note} {The BKT phase of the weakly confined floppy string is similar to the BKT phase of the 1D $p$-state quantum clock model with a global internal symmetry $\protect \mathbb {Z}_p$, which has an intermediate BKT phase with emergent $U(1)$ symmetry for $p\geq 5$~\cite {TSUI_2017}, even though the Hamiltonian has a discrete symmetry.}\BibitemShut {Stop}%
\bibitem [{\citenamefont {Wiegmann}(1978)}]{Wiegmann_1978}%
  \BibitemOpen
  \bibfield  {author} {\bibinfo {author} {\bibfnamefont {P.~B.}\ \bibnamefont {Wiegmann}},\ }\bibfield  {title} {\bibinfo {title} {One-dimensional fermi system and plane xy model},\ }\href {https://doi.org/10.1088/0022-3719/11/8/019} {\bibfield  {journal} {\bibinfo  {journal} {Journal of Physics C: Solid State Physics}\ }\textbf {\bibinfo {volume} {11}},\ \bibinfo {pages} {1583} (\bibinfo {year} {1978})}\BibitemShut {NoStop}%
\bibitem [{\citenamefont {Matsuo}\ and\ \citenamefont {Nomura}(2006)}]{Matsuo_2006}%
  \BibitemOpen
  \bibfield  {author} {\bibinfo {author} {\bibfnamefont {H.}~\bibnamefont {Matsuo}}\ and\ \bibinfo {author} {\bibfnamefont {K.}~\bibnamefont {Nomura}},\ }\bibfield  {title} {\bibinfo {title} {Berezinskii–kosterlitz–thouless transitions in the six-state clock model},\ }\href {https://doi.org/10.1088/0305-4470/39/12/006} {\bibfield  {journal} {\bibinfo  {journal} {Journal of Physics A: Mathematical and General}\ }\textbf {\bibinfo {volume} {39}},\ \bibinfo {pages} {2953} (\bibinfo {year} {2006})}\BibitemShut {NoStop}%
\bibitem [{Note2()}]{Note2}%
  \BibitemOpen
  \bibinfo {note} {When $h > h_{\protect \text {BKT}}$, the correlation length $\xi (\chi \to \infty )$ is finite. However, as long as we remain in the regime where $\xi (\chi ) \ll \xi (\chi \to \infty )$, we can still extract a scaling dimension from the finite-$\chi $ data.}\BibitemShut {Stop}%
\bibitem [{\citenamefont {Grushin}\ \emph {et~al.}(2015)\citenamefont {Grushin}, \citenamefont {Motruk}, \citenamefont {Zaletel},\ and\ \citenamefont {Pollmann}}]{Pollmann_FCI_2015}%
  \BibitemOpen
  \bibfield  {author} {\bibinfo {author} {\bibfnamefont {A.~G.}\ \bibnamefont {Grushin}}, \bibinfo {author} {\bibfnamefont {J.}~\bibnamefont {Motruk}}, \bibinfo {author} {\bibfnamefont {M.~P.}\ \bibnamefont {Zaletel}},\ and\ \bibinfo {author} {\bibfnamefont {F.}~\bibnamefont {Pollmann}},\ }\bibfield  {title} {\bibinfo {title} {Characterization and stability of a fermionic $\ensuremath{\nu}=1/3$ fractional chern insulator},\ }\href {https://doi.org/10.1103/PhysRevB.91.035136} {\bibfield  {journal} {\bibinfo  {journal} {Phys. Rev. B}\ }\textbf {\bibinfo {volume} {91}},\ \bibinfo {pages} {035136} (\bibinfo {year} {2015})}\BibitemShut {NoStop}%
\bibitem [{\citenamefont {Hauschild}\ and\ \citenamefont {Pollmann}(2018)}]{tenpy}%
  \BibitemOpen
  \bibfield  {author} {\bibinfo {author} {\bibfnamefont {J.}~\bibnamefont {Hauschild}}\ and\ \bibinfo {author} {\bibfnamefont {F.}~\bibnamefont {Pollmann}},\ }\bibfield  {title} {\bibinfo {title} {{Efficient numerical simulations with Tensor Networks: Tensor Network Python (TeNPy)}},\ }\href {https://doi.org/10.21468/SciPostPhysLectNotes.5} {\bibfield  {journal} {\bibinfo  {journal} {SciPost Phys. Lect. Notes}\ ,\ \bibinfo {pages} {5}} (\bibinfo {year} {2018})},\ \bibinfo {note} {code available from \url{https://github.com/tenpy/tenpy}},\ \Eprint {https://arxiv.org/abs/1805.00055} {arXiv:1805.00055} \BibitemShut {NoStop}%
\bibitem [{\citenamefont {Zaletel}\ \emph {et~al.}(2014)\citenamefont {Zaletel}, \citenamefont {Mong},\ and\ \citenamefont {Pollmann}}]{Zaletel_2014}%
  \BibitemOpen
  \bibfield  {author} {\bibinfo {author} {\bibfnamefont {M.~P.}\ \bibnamefont {Zaletel}}, \bibinfo {author} {\bibfnamefont {R.~S.~K.}\ \bibnamefont {Mong}},\ and\ \bibinfo {author} {\bibfnamefont {F.}~\bibnamefont {Pollmann}},\ }\bibfield  {title} {\bibinfo {title} {Flux insertion, entanglement, and quantized responses},\ }\href {https://doi.org/10.1088/1742-5468/2014/10/P10007} {\bibfield  {journal} {\bibinfo  {journal} {Journal of Statistical Mechanics: Theory and Experiment}\ }\textbf {\bibinfo {volume} {2014}},\ \bibinfo {pages} {P10007} (\bibinfo {year} {2014})}\BibitemShut {NoStop}%
\bibitem [{\citenamefont {Khemani}\ \emph {et~al.}(2016)\citenamefont {Khemani}, \citenamefont {Pollmann},\ and\ \citenamefont {Sondhi}}]{DMRG_X_2016}%
  \BibitemOpen
  \bibfield  {author} {\bibinfo {author} {\bibfnamefont {V.}~\bibnamefont {Khemani}}, \bibinfo {author} {\bibfnamefont {F.}~\bibnamefont {Pollmann}},\ and\ \bibinfo {author} {\bibfnamefont {S.~L.}\ \bibnamefont {Sondhi}},\ }\bibfield  {title} {\bibinfo {title} {Obtaining highly excited eigenstates of many-body localized hamiltonians by the density matrix renormalization group approach},\ }\href {https://doi.org/10.1103/PhysRevLett.116.247204} {\bibfield  {journal} {\bibinfo  {journal} {Phys. Rev. Lett.}\ }\textbf {\bibinfo {volume} {116}},\ \bibinfo {pages} {247204} (\bibinfo {year} {2016})}\BibitemShut {NoStop}%
\bibitem [{\citenamefont {Devakul}\ \emph {et~al.}(2017)\citenamefont {Devakul}, \citenamefont {Khemani}, \citenamefont {Pollmann}, \citenamefont {Huse},\ and\ \citenamefont {Sondhi}}]{DMRG_X_2017}%
  \BibitemOpen
  \bibfield  {author} {\bibinfo {author} {\bibfnamefont {T.}~\bibnamefont {Devakul}}, \bibinfo {author} {\bibfnamefont {V.}~\bibnamefont {Khemani}}, \bibinfo {author} {\bibfnamefont {F.}~\bibnamefont {Pollmann}}, \bibinfo {author} {\bibfnamefont {D.~A.}\ \bibnamefont {Huse}},\ and\ \bibinfo {author} {\bibfnamefont {S.~L.}\ \bibnamefont {Sondhi}},\ }\bibfield  {title} {\bibinfo {title} {Obtaining highly excited eigenstates of the localized xx chain via dmrg-x},\ }\href {https://doi.org/10.1098/rsta.2016.0431} {\bibfield  {journal} {\bibinfo  {journal} {Philosophical Transactions of the Royal Society A: Mathematical, Physical and Engineering Sciences}\ }\textbf {\bibinfo {volume} {375}},\ \bibinfo {pages} {20160431} (\bibinfo {year} {2017})}\BibitemShut {NoStop}%
\bibitem [{\citenamefont {Xu}\ \emph {et~al.}(2020)\citenamefont {Xu}, \citenamefont {Zhang},\ and\ \citenamefont {Zhang}}]{Xu_2020}%
  \BibitemOpen
  \bibfield  {author} {\bibinfo {author} {\bibfnamefont {W.-T.}\ \bibnamefont {Xu}}, \bibinfo {author} {\bibfnamefont {Q.}~\bibnamefont {Zhang}},\ and\ \bibinfo {author} {\bibfnamefont {G.-M.}\ \bibnamefont {Zhang}},\ }\bibfield  {title} {\bibinfo {title} {Tensor network approach to phase transitions of a non-abelian topological phase},\ }\href {https://doi.org/10.1103/PhysRevLett.124.130603} {\bibfield  {journal} {\bibinfo  {journal} {Phys. Rev. Lett.}\ }\textbf {\bibinfo {volume} {124}},\ \bibinfo {pages} {130603} (\bibinfo {year} {2020})}\BibitemShut {NoStop}%
\bibitem [{\citenamefont {Xu}\ and\ \citenamefont {Schuch}(2021)}]{Xu_2021}%
  \BibitemOpen
  \bibfield  {author} {\bibinfo {author} {\bibfnamefont {W.-T.}\ \bibnamefont {Xu}}\ and\ \bibinfo {author} {\bibfnamefont {N.}~\bibnamefont {Schuch}},\ }\bibfield  {title} {\bibinfo {title} {Characterization of topological phase transitions from a non-abelian topological state and its galois conjugate through condensation and confinement order parameters},\ }\href {https://doi.org/10.1103/PhysRevB.104.155119} {\bibfield  {journal} {\bibinfo  {journal} {Phys. Rev. B}\ }\textbf {\bibinfo {volume} {104}},\ \bibinfo {pages} {155119} (\bibinfo {year} {2021})}\BibitemShut {NoStop}%
\bibitem [{\citenamefont {Xu}\ \emph {et~al.}(2022)\citenamefont {Xu}, \citenamefont {Garre-Rubio},\ and\ \citenamefont {Schuch}}]{Xu_2022}%
  \BibitemOpen
  \bibfield  {author} {\bibinfo {author} {\bibfnamefont {W.-T.}\ \bibnamefont {Xu}}, \bibinfo {author} {\bibfnamefont {J.}~\bibnamefont {Garre-Rubio}},\ and\ \bibinfo {author} {\bibfnamefont {N.}~\bibnamefont {Schuch}},\ }\bibfield  {title} {\bibinfo {title} {Complete characterization of non-abelian topological phase transitions and detection of anyon splitting with projected entangled pair states},\ }\href {https://doi.org/10.1103/PhysRevB.106.205139} {\bibfield  {journal} {\bibinfo  {journal} {Phys. Rev. B}\ }\textbf {\bibinfo {volume} {106}},\ \bibinfo {pages} {205139} (\bibinfo {year} {2022})}\BibitemShut {NoStop}%
\bibitem [{\citenamefont {Borla}\ \emph {et~al.}(2025)\citenamefont {Borla}, \citenamefont {Osborne}, \citenamefont {Moroz},\ and\ \citenamefont {Halimeh}}]{borla_2025}%
  \BibitemOpen
  \bibfield  {author} {\bibinfo {author} {\bibfnamefont {U.}~\bibnamefont {Borla}}, \bibinfo {author} {\bibfnamefont {J.~J.}\ \bibnamefont {Osborne}}, \bibinfo {author} {\bibfnamefont {S.}~\bibnamefont {Moroz}},\ and\ \bibinfo {author} {\bibfnamefont {J.~C.}\ \bibnamefont {Halimeh}},\ }\href {https://arxiv.org/abs/2501.17929} {\bibinfo {title} {String breaking in a $2+1$d $\mathbb{Z}_2$ lattice gauge theory}} (\bibinfo {year} {2025}),\ \Eprint {https://arxiv.org/abs/2501.17929} {arXiv:2501.17929 [quant-ph]} \BibitemShut {NoStop}%
\bibitem [{\citenamefont {Allais}\ and\ \citenamefont {Caselle}(2009)}]{Rough_at_QCP_2009}%
  \BibitemOpen
  \bibfield  {author} {\bibinfo {author} {\bibfnamefont {A.}~\bibnamefont {Allais}}\ and\ \bibinfo {author} {\bibfnamefont {M.}~\bibnamefont {Caselle}},\ }\bibfield  {title} {\bibinfo {title} {On the linear increase of the flux tube thickness near the deconfinement transition},\ }\href {https://doi.org/10.1088/1126-6708/2009/01/073} {\bibfield  {journal} {\bibinfo  {journal} {Journal of High Energy Physics}\ }\textbf {\bibinfo {volume} {2009}},\ \bibinfo {pages} {073} (\bibinfo {year} {2009})}\BibitemShut {NoStop}%
\bibitem [{\citenamefont {Caselle}(2010)}]{Rough_at_QCP_2010}%
  \BibitemOpen
  \bibfield  {author} {\bibinfo {author} {\bibfnamefont {M.}~\bibnamefont {Caselle}},\ }\bibfield  {title} {\bibinfo {title} {Flux tube delocalization at the deconfinement point},\ }\href {https://doi.org/10.1007/JHEP08(2010)063} {\bibfield  {journal} {\bibinfo  {journal} {Journal of High Energy Physics}\ }\textbf {\bibinfo {volume} {2010}},\ \bibinfo {pages} {63} (\bibinfo {year} {2010})}\BibitemShut {NoStop}%
\bibitem [{\citenamefont {Xu}\ and\ \citenamefont {Huang}(2025)}]{Xu_huang_2024}%
  \BibitemOpen
  \bibfield  {author} {\bibinfo {author} {\bibfnamefont {W.-T.}\ \bibnamefont {Xu}}\ and\ \bibinfo {author} {\bibfnamefont {R.-Z.}\ \bibnamefont {Huang}},\ }\bibfield  {title} {\bibinfo {title} {Finite correlation length scaling of disorder parameter at quantum criticality},\ }\href {https://doi.org/10.1103/PhysRevLett.134.146503} {\bibfield  {journal} {\bibinfo  {journal} {Phys. Rev. Lett.}\ }\textbf {\bibinfo {volume} {134}},\ \bibinfo {pages} {146503} (\bibinfo {year} {2025})}\BibitemShut {NoStop}%
\bibitem [{\citenamefont {Xu}\ \emph {et~al.}(2025{\natexlab{b}})\citenamefont {Xu}, \citenamefont {Pollmann},\ and\ \citenamefont {Knap}}]{xu_FM_2024}%
  \BibitemOpen
  \bibfield  {author} {\bibinfo {author} {\bibfnamefont {W.-T.}\ \bibnamefont {Xu}}, \bibinfo {author} {\bibfnamefont {F.}~\bibnamefont {Pollmann}},\ and\ \bibinfo {author} {\bibfnamefont {M.}~\bibnamefont {Knap}},\ }\bibfield  {title} {\bibinfo {title} {Critical behavior of fredenhagen-marcu string order parameters at topological phase transitions with emergent higher-form symmetries},\ }\href {https://doi.org/10.1038/s41534-025-01030-z} {\bibfield  {journal} {\bibinfo  {journal} {npj Quantum Information}\ }\textbf {\bibinfo {volume} {11}},\ \bibinfo {pages} {74} (\bibinfo {year} {2025}{\natexlab{b}})}\BibitemShut {NoStop}%
\bibitem [{\citenamefont {Xu}\ \emph {et~al.}(2025{\natexlab{c}})\citenamefont {Xu}, \citenamefont {Knap},\ and\ \citenamefont {Pollmann}}]{zenodo}%
  \BibitemOpen
  \bibfield  {author} {\bibinfo {author} {\bibfnamefont {W.-T.}\ \bibnamefont {Xu}}, \bibinfo {author} {\bibfnamefont {M.}~\bibnamefont {Knap}},\ and\ \bibinfo {author} {\bibfnamefont {F.}~\bibnamefont {Pollmann}},\ }\bibfield  {title} {\bibinfo {title} {Tensor-network study of the roughening transition in (2 + 1)d lattice gauge theories},\ }\href {https://doi.org/10.5281/zenodo.15324471} {10.5281/zenodo.15324471} (\bibinfo {year} {2025}{\natexlab{c}})\BibitemShut {NoStop}%
\bibitem [{\citenamefont {Trebst}\ \emph {et~al.}(2007)\citenamefont {Trebst}, \citenamefont {Werner}, \citenamefont {Troyer}, \citenamefont {Shtengel},\ and\ \citenamefont {Nayak}}]{Trebst_2007}%
  \BibitemOpen
  \bibfield  {author} {\bibinfo {author} {\bibfnamefont {S.}~\bibnamefont {Trebst}}, \bibinfo {author} {\bibfnamefont {P.}~\bibnamefont {Werner}}, \bibinfo {author} {\bibfnamefont {M.}~\bibnamefont {Troyer}}, \bibinfo {author} {\bibfnamefont {K.}~\bibnamefont {Shtengel}},\ and\ \bibinfo {author} {\bibfnamefont {C.}~\bibnamefont {Nayak}},\ }\bibfield  {title} {\bibinfo {title} {Breakdown of a topological phase: Quantum phase transition in a loop gas model with tension},\ }\href {https://doi.org/10.1103/PhysRevLett.98.070602} {\bibfield  {journal} {\bibinfo  {journal} {Phys. Rev. Lett.}\ }\textbf {\bibinfo {volume} {98}},\ \bibinfo {pages} {070602} (\bibinfo {year} {2007})}\BibitemShut {NoStop}%
\bibitem [{\citenamefont {Lootens}\ \emph {et~al.}(2023)\citenamefont {Lootens}, \citenamefont {Delcamp}, \citenamefont {Ortiz},\ and\ \citenamefont {Verstraete}}]{Duality_TN}%
  \BibitemOpen
  \bibfield  {author} {\bibinfo {author} {\bibfnamefont {L.}~\bibnamefont {Lootens}}, \bibinfo {author} {\bibfnamefont {C.}~\bibnamefont {Delcamp}}, \bibinfo {author} {\bibfnamefont {G.}~\bibnamefont {Ortiz}},\ and\ \bibinfo {author} {\bibfnamefont {F.}~\bibnamefont {Verstraete}},\ }\bibfield  {title} {\bibinfo {title} {Dualities in one-dimensional quantum lattice models: Symmetric hamiltonians and matrix product operator intertwiners},\ }\href {https://doi.org/10.1103/PRXQuantum.4.020357} {\bibfield  {journal} {\bibinfo  {journal} {PRX Quantum}\ }\textbf {\bibinfo {volume} {4}},\ \bibinfo {pages} {020357} (\bibinfo {year} {2023})}\BibitemShut {NoStop}%
\bibitem [{\citenamefont {Fradkin}(2017)}]{Fradkin_2017_disorder_para}%
  \BibitemOpen
  \bibfield  {author} {\bibinfo {author} {\bibfnamefont {E.}~\bibnamefont {Fradkin}},\ }\bibfield  {title} {\bibinfo {title} {Disorder operators and their descendants},\ }\href {https://doi.org/10.1007/s10955-017-1737-7} {\bibfield  {journal} {\bibinfo  {journal} {Journal of Statistical Physics}\ }\textbf {\bibinfo {volume} {167}},\ \bibinfo {pages} {427} (\bibinfo {year} {2017})}\BibitemShut {NoStop}%
\bibitem [{\citenamefont {Philippe Di~Francesco}(1997)}]{yellow_book_CFT}%
  \BibitemOpen
  \bibfield  {author} {\bibinfo {author} {\bibfnamefont {D.~S.}\ \bibnamefont {Philippe Di~Francesco}, \bibfnamefont {Pierre~Mathieu}},\ }\href@noop {} {\emph {\bibinfo {title} {Conformal Field Theory}}}\ (\bibinfo  {publisher} {Springer-Verlag New York, Inc},\ \bibinfo {year} {1997})\BibitemShut {NoStop}%
\bibitem [{\citenamefont {Bl\"ote}\ and\ \citenamefont {Deng}(2002)}]{Deng_2002}%
  \BibitemOpen
  \bibfield  {author} {\bibinfo {author} {\bibfnamefont {H.~W.~J.}\ \bibnamefont {Bl\"ote}}\ and\ \bibinfo {author} {\bibfnamefont {Y.}~\bibnamefont {Deng}},\ }\bibfield  {title} {\bibinfo {title} {Cluster monte carlo simulation of the transverse ising model},\ }\href {https://doi.org/10.1103/PhysRevE.66.066110} {\bibfield  {journal} {\bibinfo  {journal} {Phys. Rev. E}\ }\textbf {\bibinfo {volume} {66}},\ \bibinfo {pages} {066110} (\bibinfo {year} {2002})}\BibitemShut {NoStop}%
\bibitem [{\citenamefont {Tsui}\ \emph {et~al.}(2017)\citenamefont {Tsui}, \citenamefont {Huang}, \citenamefont {Jiang},\ and\ \citenamefont {Lee}}]{TSUI_2017}%
  \BibitemOpen
  \bibfield  {author} {\bibinfo {author} {\bibfnamefont {L.}~\bibnamefont {Tsui}}, \bibinfo {author} {\bibfnamefont {Y.-T.}\ \bibnamefont {Huang}}, \bibinfo {author} {\bibfnamefont {H.-C.}\ \bibnamefont {Jiang}},\ and\ \bibinfo {author} {\bibfnamefont {D.-H.}\ \bibnamefont {Lee}},\ }\bibfield  {title} {\bibinfo {title} {The phase transitions between zn×zn bosonic topological phases in 1+1d, and a constraint on the central charge for the critical points between bosonic symmetry protected topological phases},\ }\href {https://doi.org/https://doi.org/10.1016/j.nuclphysb.2017.03.021} {\bibfield  {journal} {\bibinfo  {journal} {Nuclear Physics B}\ }\textbf {\bibinfo {volume} {919}},\ \bibinfo {pages} {470} (\bibinfo {year} {2017})}\BibitemShut {NoStop}%
\end{thebibliography}%
